\documentstyle[epsf,aps,psfig]{revtex}
\renewcommand{\thefootnote}{\fnsymbol{footnote}}
\begin{document}
\setcounter{footnote}{1}
\begin{center}
{\Large\bf Debye screening and Meissner effect in a two-flavor 
color superconductor}
\\[1cm]
Dirk H.\ Rischke 
\\ ~~ \\
{\it RIKEN-BNL Research Center} \\
{\it Brookhaven National Laboratory, Upton, New York 11973, U.S.A.} \\
{\it email: rischke@bnl.gov}
\\ ~~ \\ ~~ \\
\end{center}
\begin{abstract} 
I compute the gluon self-energy in a color superconductor with
two flavors of massless quarks, where
condensation of Cooper pairs breaks $SU(3)_c$ to $SU(2)_c$.
At zero temperature, there is neither Debye screening nor a Meissner effect 
for the three gluons of the unbroken $SU(2)_c$ subgroup. The remaining five 
gluons attain an electric as well as a magnetic mass.
For temperatures approaching the critical
temperature for the onset of color superconductivity, or for
gluon momenta much larger than the color-superconducting
gap, the self-energy assumes the
form given by the standard hard-dense loop approximation.
The gluon self-energy determines the coefficient of the kinetic term
in the effective low-energy theory for the condensate fields.
\end{abstract}
\renewcommand{\thefootnote}{\arabic{footnote}}
\setcounter{footnote}{0}

\section{Introduction}

Single-gluon exchange between two quarks
is attractive in the color anti-triplet channel.
Therefore, sufficiently cold and dense quark matter is a color 
superconductor \cite{bailinlove,general}. 

In some aspects, color superconductivity is 
similar to ordinary (BCS) superconductivity \cite{BCS,fetter}. 
For instance, like electrons in a BCS superconductor, quarks form
Cooper pairs. At zero temperature, $T=0$, 
the ground state of the system is no longer
a Fermi sea of quarks (and a Dirac sea of antiquarks), but
a Bose condensate of quark Cooper pairs. In the normal phase
the excitation of a particle--hole pair at the Fermi surface costs no 
energy. In the superconducting phase, however, exciting a
quasiparticle--quasiparticle-hole pair costs at least an energy $2\, \phi_0$,
where $\phi_0$ is the zero-temperature gap.
Another similarity between color and BCS superconductivity is
that, in weak coupling, the critical temperature $T_c$ for ``melting'' the
Cooper pair condensate is $T_c \simeq 0.57 \, \phi_0$ \cite{prlett2,prqcd}.

There are, however, also fundamental differences between
color and BCS superconductivity.
First of all, a BCS superconductor requires the presence of
an atomic lattice with phonons
that cause electrons to form Cooper pairs. On the other hand, 
in QCD gluons themselves cause quarks to condense.
Another difference is that in BCS theory
the zero-temperature gap depends on the BCS coupling constant $G$ as
$\phi_0 \sim \mu \, \exp(-c_{\rm BCS}/G^2)$ \cite{BCS,fetter},
where $\mu$ is the chemical potential, and $c_{\rm BCS} = $const., while
in a color superconductor,
$\phi_0 \sim \mu \, \exp(-c_{\rm QCD}/g)$ \cite{prlett,Son},
where $g$ is the QCD coupling constant, and $c_{\rm BCS} \neq c_{\rm QCD}=$ 
const.

The physical reason for the change in the parametric dependence on the 
coupling constant is that, because gluons are massless, gluon-mediated
interactions are long-range, in contrast to BCS theory, where
phonon exchange is typically assumed to be a point-like interaction
\cite{BCS,fetter}. The long-range nature of gluon exchange manifests itself 
in the infrared singular behavior of the gluon propagator. This
enhances the contribution of very soft, collinear gluons
in the gap equations \cite{prlett2,prqcd}, and causes the $1/g$ in 
the exponent, instead of a $1/g^2$ which would appear if 
gluons were massive \cite{prscalar}, or gluon exchange a
point-like interaction as assumed in Nambu--Jona-Lasinio-type
approaches to color superconductivity \cite{general}.

Some care has to be taken in determining the coefficient
$c_{\rm QCD}$. This constant differs when one uses 
the free gluon propagator \cite{hong} in the solution of the
gap equations instead of a propagator which takes into account the
presence of the cold and dense quark medium.
By now, several authors 
\cite{prlett2,prqcd,hongetal,SchaferWilczek,rockefeller,hsuschwetz}
have confirmed Son's original result $c_{\rm QCD}
= 3 \pi^2/\sqrt{2}$ \cite{Son}, obtained by using the gluon propagator in
the so-called ``hard-dense-loop'' (HDL) limit
\cite{LeBellac,finitedens}. The gluon propagator in the HDL limit is 
obtained by resummation of the gluon self-energy, computed to one-loop order
for gluon energies $p_0$ and momenta $p$ that are much smaller
than the quark chemical potential $\mu$.

In weak coupling, the temperatures where quark matter
is color-superconducting are much smaller than
the quark chemical potential, $T \sim \phi_0 \sim \mu \,
\exp( - c_{\rm QCD}/g) \ll \mu$. Therefore, to leading order
the contributions of gluon and ghost loops to
the one-loop gluon self-energy can be neglected, and the
main contribution comes from the quark loop.
This is very similar to ordinary superconductivity, where the
one-loop photon self-energy is determined by an electron loop.

In the standard HDL approximation, however, the
quark excitations in the loop are considered to be those of the normal and
not of the superconducting phase. This is in principle inconsistent.
The aim of the present work is to amend this
shortcoming and to compute the gluon self-energy 
in the color-superconducting phase.

For the sake of definiteness, I consider a color superconductor
with $N_f = 2$ flavors of massless quarks, and assume that quarks condense 
in a channel with total spin $J=0$ and even parity. In this case,
the quark-quark condensate breaks $SU(3)_c$ to $SU(2)_c$.
Consequently, one expects that
the three gluons of the unbroken $SU(2)_c$ subgroup remain
massless, while the other five gluons of the original $SU(3)_c$
obtain masses through the Anderson--Higgs mechanism. It is therefore
necessary to consider different gluon colors separately.

I derive a general expression for the quark contribution to
the gluon self-energy, and study the limit where the gluon energy
$p_0 = 0$ and the gluon momentum $p \rightarrow 0$.
For electric gluons, this limit gives the Debye mass, while for
magnetic gluons, it gives the Meissner mass.
I also consider the limit where $p_0 = 0$, but $p \gg \phi_0$.
In this case, the gluon momentum is large enough to resolve individual 
quarks in a Cooper pair; consequently,
the Debye masses approach their values in the normal phase and
the Meissner effect vanishes.

Debye screening of static color-electric fields
and the Meissner effect for static color-magnetic
fields are in principle quite analogous to Debye screening 
and the Meissner effect for electromagnetic fields
in ordinary superconductors \cite{BCS,fetter}. 
However, the somewhat more complicated color and flavor 
structure of a quark-quark condensate in comparison to an electron-electron 
condensate gives rise to an additional degree of
complexity. While studying these effects in a color superconductor is
interesting in itself, they might have, however, far greater implications
for color superconductivity than the corresponding effects in
ordinary superconductors:
unlike photons, gluons themselves are responsible for condensation
of quark pairs. The modification of the gluon self-energy
in the superconducting phase directly enters the gap equation
through the gluon propagator, and so might change the value for
the gap. On the other hand, the influence of the photon self-energy on 
electron condensation is at best a higher order effect.

Although effects from quark condensation in the gluon
propagator vanish for large
gluon energies and momenta, one can {\em a priori\/} not
exclude that they will not change the solution of the gap equations.
For instance, to assess the importance of the Meissner effect,
note that, in the HDL approximation, the main contribution to the 
gap equations comes from color-magnetic fields with
momenta $p \sim (m_g^2\, \phi_0)^{1/3} \gg \phi_0$, where
$m_g$ is the gluon mass \cite{prlett2,prqcd,Son,SchaferWilczek}. 
As will be seen below, the Meissner effect
is small, but not absent, at the same momentum scale. 
This means that the Meissner effect can indeed influence the
solution of the gap equation. A first estimate of this
effect (neglecting the color-flavor structure of the condensate
and considering only the dominant contribution to the gluon
self-energy) was given in \cite{ehhs}, and a 
reduction of the zero-temperature gap was found.

This paper is organized as follows.
In Sec.\ \ref{ii} a compact derivation of the quark contribution
to the gluon self-energy is presented, mainly to introduce 
the notation and the concept of Nambu--Gor'kov spinors 
\cite{fetter}, which considerably simplify calculations at nonzero chemical
potential. In Sec.\ \ref{iii} the quark contribution to the gluon
self-energy is explicitly computed in the normal phase.
The HDL limit is derived
to show that the Nambu--Gor'kov method indeed gives the correct answer. 
Section \ref{iv} generalizes the previous results to the superconducting
phase. In Sec.\ \ref{v}, the zero-energy, zero-momentum limit
of the gluon self-energy is studied, which yields the Debye as well as
the Meissner masses in the superconducting phase. 
Section \ref{vi} discusses how, for nonzero gluon momenta $p \gg \phi_0$,
the Debye masses approach their values in the normal phase, and
the Meissner effect vanishes.
Readers not interested in technical details should 
skip Secs.\ \ref{ii} to \ref{vi} and move on to Sec.\ \ref{vii}, where
the main results of this work are summarized, conclusions are drawn, and 
an outlook for future studies is given.

I use natural units, $\hbar=c=k_B=1$, and
work in Euclidean space-time ${\bf R}^4 \equiv V/T$, where $V$ is the volume
and $T$ the temperature of the system. Nevertheless, I find it convenient to
retain the Minkowski notation for 4-vectors, with a metric tensor
$g^{\mu \nu} = {\rm diag}(+,-,-,-)$. For instance, the space-time
coordinate vector is $x^\mu \equiv (t,{\bf x})$, $t \equiv -i\tau$, where
$\tau$ is Euclidean time. 4-momenta are denoted as
$K^\mu \equiv (k_0,{\bf k})$, $k_0 \equiv -i \omega_n$, where
$\omega_n$ is the Matsubara frequency, $\omega_n \equiv 2n \pi T$
for bosons and $\omega_n \equiv (2n+1) \pi T$ for fermions, $n=0,\pm 1,\pm 2,
\ldots$. The absolute value of the 3-momentum ${\bf k}$ is denoted as
$k \equiv |{\bf k}|$, and its direction as $\hat{\bf k} \equiv {\bf k}/k$.

\section{The generating functional at nonzero chemical potential} \label{ii}

Consider QCD with $N_f$ quark flavors, at nonzero chemical potential.
The $N_f \times N_f$ matrix of quark masses $m_f$ will be denoted as
$m \equiv {\rm diag}(m_1,m_2,\ldots,m_{N_f})$. Let us consider
a color neutral system, {\it i.e.}, there is no chemical potential
for color, however, there can be in general
a chemical potential $\mu_f$ for each quark flavor
$f$. Let us denote the $N_f \times N_f$ chemical potential matrix
as $\mu \equiv {\rm diag}(\mu_1,\mu_2,\ldots,\mu_{N_f})$.
Then, the generating functional for the $N$--point functions of
the theory reads (normalization factors are suppressed)
\begin{mathletters}
\begin{eqnarray} \label{genfunc}
{\cal Z}[J,\bar{\eta},\eta] & = & 
\int {\cal D} U[A] \; \exp\left[ \int_x \left( {\cal L}_A + J_\mu^a A_a^\mu
\right) \right] \;{\cal Z}[A,\bar{\eta},\eta]\,\, ,  \\
{\cal Z}[A,\bar{\eta},\eta] & = & 
\int {\cal D} \bar{\psi} \, {\cal D} \psi\,
\exp \left\{ \int_x  \left[ \bar{\psi} \left(
i \gamma^\mu \partial_\mu + \mu \gamma_0 -m +g \gamma^\mu A_\mu^a T_a
\right) \psi + \bar{\eta} \psi + \bar{\psi} \eta \right] \right\}
\,. \label{genfuncquarks}
\end{eqnarray}
\end{mathletters}
Here, ${\cal D}U[A]$ is the
gauge invariant measure for the integration over the gauge fields
$A_a^\mu$. The space-time integration is defined as
$\int_x \equiv \int_0^{1/T} d\tau \int_V d^3{\bf x}\,$. $g$ is the QCD
coupling constant, $\gamma^\mu$ are the Dirac matrices, and $T_a= \lambda_a/2$ 
the generators of $SU(N_c)$; for QCD, $N_c = 3$, and $\lambda_a$ are the
Gell-Mann matrices. The quark fields $\psi$ (as well as the external
fields $\eta$) are $4 N_c N_f$-component spinors, {\it i.e.}, 
they carry Dirac indices $\alpha = 1, \ldots,4$, fundamental color indices
$i=1,\ldots,N_c$, and flavor indices $f=1,\ldots,N_f$.
The Lagrangian for the gauge fields consists in general
of three parts,
\begin{equation}
{\cal L}_A = {\cal L}_F + {\cal L}_{\rm gf} + {\cal L}_{FPG}\,\, ,
\end{equation}
where 
\begin{equation}
{\cal L}_F = - \frac{1}{4} F^{\mu \nu}_a F_{\mu \nu}^a
\end{equation}
is the gauge field part, $F_{\mu\nu}^a = \partial_\mu A_\nu^a 
- \partial_\nu A_\mu^a + g f^{abc} A_\mu^b A_\nu^c$ is the
field strength tensor. The parts corresponding to gauge fixing,
${\cal L}_{\rm gf}$, and to Fadeev--Popov ghosts,
${\cal L}_{FPG}$, need not be specified: it will be seen that they
are inconsequential for the following.

In the vacuum, the ground state of the system consists of
the Dirac sea, {\it i.e.}, all negative energy (antiquark) states
are occupied, while all positive energy (quark) states are empty.
At zero temperature and nonzero chemical potential, $\mu_f>0$, however,
the ground state consists of the Dirac sea {\em and\/} the
Fermi sea, {\it i.e.}, positive energy states which are occupied up to the 
Fermi energy $\mu_f$. 
Formally, this is expressed by the term $\bar{\psi} \mu \gamma_0 \psi$ 
in the generating functional (\ref{genfuncquarks}), which ensures
that the energy of excited states of flavor $f$ is measured with respect
to the Fermi energy $\mu_f$, and not with respect to the vacuum at
zero density. 

This shift of the energy scale introduces an apparent asymmetry. 
One can restore the symmetry by the following trick.
Introduce $M$ identical copies (``replicas'') of the original
quark fields. All copies are supposed to interact with the
gluon field in the same way. At the end, after having computed 
$N$-point functions for this extended system, $M$ will be set equal to 1.
The generating functional (\ref{genfuncquarks}) for the quark
part is replaced by
\begin{equation} \label{genfuncquarksM}
{\cal Z} [A,\bar{\eta},\eta]  \rightarrow
{\cal Z}_M [A,\bar{\eta},\eta] \equiv  
\left({\cal Z}[A,\bar{\eta},\eta] \right)^M \,\, .
\end{equation}
Now define the charge conjugate spinors $\psi_C,\, \bar{\psi}_C$ 
through
\begin{equation} \label{chargeconj}
\psi \equiv C\, \bar{\psi}_C^T \;\;\;\; , \;\;\;\; \bar{\psi} \equiv
\psi_C^T\, C\,\, ,
\end{equation}
where $C \equiv i \gamma^2 \gamma_0$ is the charge conjugation matrix;
$C = -C^{-1} = -C^T = -C^\dagger$. In half of the $M$ copies
in Eq.\ (\ref{genfuncquarksM}),
replace $\bar{\psi}$, $\psi$ by the charge conjugate spinors
$\bar{\psi}_C$, $\psi_C$.
Using $C \gamma_\mu C^{-1} = - \gamma_\mu^T$, and the anticommutation
property of the (Grassmann-valued) quark spinors, one obtains after
an integration by parts (and disregarding the overall normalization)
\begin{eqnarray}
{\cal Z}_M[A,\bar{\eta},\eta,\bar{\eta}_C,\eta_C]
& = & \left(\int {\cal D} \bar{\psi} \, {\cal D} \psi \,
{\cal D} \bar{\psi}_C \, {\cal D} \psi_C \; \exp \left\{ \int_x  \left[
\bar{\psi} \left( i \gamma^\mu \partial_\mu + \mu \gamma_0 -m +g
A_\mu^a \Gamma^\mu_a \right) \psi  \right. \right. \right. \nonumber \\
&        & \left. \left. \left. + \; 
\bar{\psi}_C \left( i \gamma^\mu \partial_\mu - \mu \gamma_0 -m + g
A_\mu^a \bar{\Gamma}^\mu_a \right) \psi_C + \bar{\eta} \psi 
+\bar{\psi} \eta +\bar{\eta}_C \psi_C
+ \bar{\psi}_C \eta_C \right] \frac{}{} \right\} \right)^{M/2}\,\, .
\label{genfuncquarks2}
\end{eqnarray} 
Here,
\begin{equation}
\Gamma^\mu_a \equiv \gamma^\mu T_a \;\;\;\; , \;\;\;\;
\bar{\Gamma}^\mu_a \equiv C (\gamma^\mu)^T C^{-1} T_a^T \equiv -\gamma^\mu
T_a^T \,\, ,
\end{equation}
and charge conjugate external fields
$\bar{\eta}_C$ and $\eta_C$ were defined analogous to Eq.\
(\ref{chargeconj}).
Let us now introduce the $8 N_c N_f$-component (Nambu--Gor'kov) spinors
\begin{equation}
\Psi \equiv \left( \begin{array}{c} 
                    \psi \\
                    \psi_C 
                   \end{array}
            \right) \,\,\, , \,\,\,\,
\bar{\Psi} \equiv ( \bar{\psi} \, , \, \bar{\psi}_C ) 
\,\,\, , \,\,\,\,
H \equiv \left( \begin{array}{c} 
                    \eta \\
                    \eta_C 
                   \end{array}
            \right) \,\,\, , \,\,\,\,
\bar{H} \equiv ( \bar{\eta} \, , \, \bar{\eta}_C ) 
\,\, , 
\end{equation}
and the $8 N_c N_f \times 8 N_c N_f$-dimensional inverse propagator 
\begin{equation} \label{S0-1}
{\cal S}_0^{-1}(x,y) \equiv
\left( \begin{array}{cc}
            [G_0^+]^{-1}(x,y) & 0 \\
             0 & [G_0^-]^{-1}(x,y) 
       \end{array} \right)\,\, ,
\end{equation}
where
\begin{equation} \label{G0pm-1}
[G_0^\pm]^{-1}(x,y) \equiv -i \left( i \gamma_\mu \partial^\mu_x
\pm \mu \gamma_0 -m \right) \delta^{(4)}(x-y) 
\end{equation}
is the inverse propagator for non-interacting quarks (upper sign) or 
charge conjugate quarks (lower sign), respectively.
Furthermore, denote
\begin{equation} \label{Gamma}
\hat{\Gamma}^\mu_a \equiv \left( \begin{array}{cc}
                                 \Gamma^\mu_a & 0 \\
                                 0 & \bar{\Gamma}^\mu_a
                          \end{array} \right)\,\, .
\end{equation}
Then, the generating functional (\ref{genfuncquarks2}) can be
written in the compact form
\begin{eqnarray} 
\lefteqn{{\cal Z}_M[A,\bar{H},H] } \nonumber \\
& = & 
\int \prod_{r=1}^{M/2} {\cal D} \bar{\Psi}_r \, {\cal D} \Psi_r \; \exp 
\left\{ \sum_{r=1}^{M/2} \left[
\int_{x,y} \bar{\Psi}_r(x) \,{\cal S}_0^{-1} (x,y)\,
\Psi_r(y) + \int_x \left( g\, \bar{\Psi}_r \,  A_\mu^a\,\hat{\Gamma}^\mu_a\,
\Psi_r + \bar{H}_r \Psi_r + \bar{\Psi}_r H_r \right) \right] \right\}
\,\, . \label{genfuncquarks3}
\end{eqnarray}
In this form, all reference to the chemical potentials $\mu_f$
has been absorbed in the inverse propagator (\ref{S0-1}).
Therefore, the generating functional for QCD, Eq.\ (\ref{genfunc}) with
(\ref{genfuncquarks3}), is formally identical to that at zero chemical
potential. The apparent asymmetry introduced by a nonzero chemical potential
$\mu_f$ has been restored by the introduction of charge conjugate fields;
the associated charge conjugate propagator $G_0^-$ appears on equal
footing with the ordinary propagator $G_0^+$.

\section{The gluon self-energy in the normal phase}\label{iii}

The gluon self-energy is defined as
\begin{equation}
\Pi \equiv 
\Delta^{-1} - \Delta_0^{-1}\,\,,
\end{equation}
where $\Delta^{-1}$ is the resummed and
$\Delta_0^{-1}$ the free inverse gluon propagator; for instance,
in momentum space and in covariant gauge,
\begin{equation}
[\Delta_0^{-1}]^{\mu \nu}_{ab}(P) = \delta_{ab}\, \left( P^2 g^{\mu \nu}
+ \frac{1-\alpha}{\alpha}\, P^\mu P^\nu \right)\,\,.
\end{equation}
To one-loop order, the gluon self-energy receives contributions from
gluon loops (through the 3-gluon and 4-gluon vertices), 
ghost loops (through the ghost-gluon vertex), and quark loops
(through the quark-gluon vertex),
\begin{equation}
\Pi = \Pi_g + \Pi_{FPG} + \Pi_q + O(g^3)\,\,.
\end{equation}
$\Pi_g$ and $\Pi_{FPG}$ are independent of $\mu$, effects from nonzero 
chemical potential enter only through $\Pi_q$. For dimensional reasons,
\begin{equation}
\Pi_g\, , \; \Pi_{FPG} \sim g^2 T^2 \;\;\; , \;\;\;\; 
\Pi_q \sim g^2 (\mu^2 + a\, T^2)\,\, ,
\end{equation}
with some constant $a$.

The superconducting condensate melts when the temperature $T$
exceeds the critical temperature $T_c \simeq 0.57 \, \phi_0$ 
\cite{prlett2,prqcd}, where
$\phi_0$ is the magnitude of the superconducting gap at $T=0$.
In weak coupling QCD, $\phi_0 \sim \mu \, \exp(-c_{\rm QCD}/g) \ll \mu$ 
\cite{prlett2,prqcd,prlett,Son,hongetal,SchaferWilczek,rockefeller,hsuschwetz},
and temperature effects can be neglected to leading order. 
This means that, for the temperatures of interest in this work, one can
neglect the contributions from gluon and ghost loops to the
gluon self-energy, and consider the quark contribution only,
$\Pi \simeq \Pi_q$.

Due to the aforementioned similarity between the generating functional
(\ref{genfunc}), with the quark part (\ref{genfuncquarks3}), 
and the one at zero chemical potential, it is not difficult to derive the 
quark contribution to the one-loop gluon self-energy. 
If there is no superconducting condensate, this contribution is
\begin{equation} \label{Pixy1}
{\Pi_0}^{\mu \nu}_{ab} (x,y) \equiv 
\frac{M}{2} \,g^2\, {\rm Tr}_{s,c,f,NG} \left[ \hat{\Gamma}^\mu_a
\, {\cal S}_0 (x,y) \, \hat{\Gamma}^\nu_b \, {\cal S}_0(y,x) \right]
\,\,.
\end{equation}
Here, the factor $M/2$ arises from the fact that there are
$M/2$ identical species of quarks described by spinors $\Psi_r$
in Eq.\ (\ref{genfuncquarks3}), which contribute to the gluon self-energy.
In the following, set $M=1$, to recover the original
theory. The trace in Eq.\ (\ref{Pixy1}) is taken over $4$-dimensional spinor 
space, $N_c$-dimensional color space, $N_f$-dimensional flavor space, and 
the $2$-dimensional space of regular and charge-conjugate
spinors (Nambu--Gor'kov space). 

In the following, the self-energy (\ref{Pixy1}) is evaluated in momentum space.
Use will be made of translational invariance, 
${\cal S}_0(x,y) \equiv {\cal S}_0(x-y)$, cf.\ Eq.\ 
(\ref{G0pm-1}), and of the Fourier transforms
\begin{mathletters}
\begin{eqnarray}
{\cal S}_0(x) & = & \frac{T}{V} \sum_K e^{-iK \cdot x}\, {\cal S}_0(K)\,\, , \\
-i\,\delta^{(4)}(x) & \equiv & \delta^{(3)}({\bf x}) \, \delta(\tau)
= \frac{T}{V} \sum_K e^{-iK \cdot x} \,\, , \\
\int_x e^{i K \cdot x} & = & \frac{V}{T}\, \delta^{(4)}_{K,0}\,\, ,
\end{eqnarray}
\end{mathletters}
where $\sum_K \equiv \sum_n V \int d^3{\bf k}/(2 \pi)^3$.
Here, the quark propagator in momentum space is
\begin{equation} \label{S0K}
{\cal S}_0(K) \equiv \left( \begin{array}{cc}
            G_0^+(K) & 0 \\
             0 & G_0^-(K) 
       \end{array} \right)\;\;\; ,\;\;\;\;
G_0^\pm(K) \equiv (\gamma^\mu K_\mu \pm \mu \gamma_0 - m)^{-1}\,\,.
\end{equation}
Then, the gluon self-energy in momentum space is
\begin{equation} \label{PiP}
{\Pi_0}^{\mu \nu}_{ab} (P) = \frac{1}{2} \,g^2 \, \frac{T}{V} 
\sum_K {\rm Tr}_{s,c,f,NG} \left[ \hat{\Gamma}^\mu_a
\, {\cal S}_0 (K) \, \hat{\Gamma}^\nu_b \, {\cal S}_0(K-P) \right]
\,\,.
\end{equation}
As a warm-up exercise, and to confirm that the method
of the Nambu--Gor'kov propagators indeed gives the correct answer,
let us derive from Eq.\ (\ref{PiP})
the standard hard-dense-loop
(HDL) result \cite{LeBellac,finitedens} 
for the quark contribution to the gluon self-energy.
To see the analogy to the computation in the superconducting
phase, cf.\ Sec.\ \ref{iv}, Eq.\ (\ref{PiP})
will be evaluated in several steps.

\subsection{Trace over Nambu--Gor'kov space}

First perform the trace over Nambu--Gor'kov space. 
With Eqs.\ (\ref{Gamma}) and (\ref{S0K}), one obtains
\begin{equation}
{\Pi_0}^{\mu \nu}_{ab} (P) = \frac{1}{2} \,g^2\,  \frac{T}{V}
\sum_K {\rm Tr}_{s,c,f} \left[\Gamma^\mu_a \, G_0^+ (K)\, \Gamma^\nu_b \,
G_0^+(K-P) + \bar{\Gamma}^\mu_a \,G_0^-(K)\,\bar{\Gamma}^\nu_b\,G_0^-(K-P)
\right]\,\, .
\end{equation}

\subsection{Trace over flavor space}

The vertices $\Gamma^\mu_a$ and $\bar{\Gamma}^\mu_a$ 
are diagonal in flavor space,
\begin{equation}
\left(\Gamma^\mu_a\right)_{fg} = \delta_{fg}\, \Gamma^\mu_a \;\;\; , \;\;\;
\left(\bar{\Gamma}^\mu_a\right)_{fg} = \delta_{fg} \, \bar{\Gamma}^\mu_a
\,\,.
\end{equation}
The free propagators $G_0^\pm$ are also diagonal in flavor space, but
for $\mu_f \neq \mu_g$, $f \neq g$, $f,g \in \{ 1,\ldots,N_f\}$, 
the diagonal components are in general not equal. 
To proceed, assume that all chemical potentials are equal,
$\mu_1 = \mu_2 = \ldots = \mu_{N_f} \equiv \mu$, such that
\begin{equation}
\left(G_0^\pm \right)_{fg} = \delta_{fg} \, G_0^\pm \,\, .
\end{equation}
(For notational convenience, I am somewhat sloppy with
indices here and throughout the rest of the paper:
I use the same symbol, $G_0^\pm$, for the
$4\, N_c\, N_f \times 4\, N_c\, N_f$ matrix on the left-hand side
of this equation and for the $4\, N_c \times 4\, N_c$ matrix on the
right-hand side.)
Thus, the trace over flavor space simply gives a factor $N_f$,
\begin{equation}
{\Pi_0}^{\mu \nu}_{ab} (P) = \frac{1}{2} \,g^2\, N_f\, \frac{T}{V}
\sum_K {\rm Tr}_{s,c} \left[\Gamma^\mu_a \, G_0^+ (K)\, \Gamma^\nu_b \,
G_0^+(K-P) + \bar{\Gamma}^\mu_a \,G_0^-(K)\,\bar{\Gamma}^\nu_b\,G_0^-(K-P)
\right]\,\, .
\end{equation}
This expression is easily generalized to the case where
the chemical potentials are not equal. Then, instead of the prefactor $N_f$
one would have a sum over flavors $f$, where the value of
the chemical potential in the propagators 
$G_0^\pm$ in the $f$th term of the sum is equal to $\mu_f$.

\subsection{Trace over color space}

The free quark propagator is diagonal in (fundamental) color space,
\begin{equation}
\left(G_0^\pm \right)_{ij} = \delta_{ij} \, G_0^\pm \,\, .
\end{equation}
The only nontrivial color structure thus arises from the
generators of $SU(3)_c$. On account of
\begin{equation}
{\rm Tr}_c(T_a T_b) = {\rm Tr}_c(T_a T_b)^T 
={\rm Tr}_c(T_a^T T_b^T) = \frac{1}{2} \, \delta_{ab} \,\,,
\end{equation} 
one obtains
\begin{mathletters}\label{Pis}
\begin{eqnarray} 
{\Pi_0}^{\mu \nu}_{ab}(P) & = & \delta_{ab}\, {\Pi_0}^{\mu \nu} (P)\,\, ,\\
{\Pi_0}^{\mu \nu} (P) & = & \frac{1}{4}\,  g^2\, N_f
\, \frac{T}{V} \sum_K {\rm Tr}_{s} 
\left[\gamma^\mu \, G_0^+ (K)\, \gamma^\nu \,
G_0^+(K-P) + \gamma^\mu \,G_0^-(K)\,\gamma^\nu \,G_0^-(K-P) \right]\,\, .
\label{Pisb}
\end{eqnarray}
\end{mathletters}

\subsection{Mixed representations for the quark propagators}

To proceed, let us assume that the quarks are massless, $m=0$. Then,
write the quark propagator as
\begin{equation} \label{freequarkprop}
G_0^\pm(K) = \sum_{e=\pm} 
\frac{k_0 \mp (\mu - ek)}{k_0^2 - [\epsilon^e_{{\bf k}0}]^2} \,
\Lambda^{\pm e}_{\bf k}\,\gamma_0 \,\, ,
\end{equation}
where
\begin{equation} \label{freequarkenergy}
\epsilon^e_{{\bf k}0} \equiv |\mu - ek|\,\, ,
\end{equation}
and
\begin{equation}
\Lambda^e_{\bf k} \equiv \frac{1}{2} \left( 1 + e \gamma_0\,
\bbox{\gamma} \cdot \hat{\bf k} \right)
\end{equation}
are projectors onto states of positive ($e = +$) or negative
($e=-1$) energy.
Now introduce a mixed representation for the quark propagator,
\begin{equation}
G_0^\pm(\tau,{\bf k}) \equiv T \sum_{k_0} e^{-k_0\tau} \,G_0^\pm(K)
\;\;\;,\;\;\;\;
G_0^\pm(K) \equiv \int_0^{1/T} d\tau\, e^{k_0\tau}\, 
G_0^\pm(\tau,{\bf k}) \,\, .
\end{equation}
After performing the Matsubara sum in terms of a contour integral in
the complex $k_0$ plane, one obtains
\begin{mathletters} \label{mixedprops}
\begin{eqnarray}
G_0^+(\tau,{\bf k}) & = & - \sum_{e = \pm} 
\Lambda^{e}_{\bf k}\, \gamma_0 \left\{ \frac{}{}\! (1-n_{{\bf k}0}^e)\,
\left[ \theta(\tau) - N(\epsilon^e_{{\bf k}0}) \right]
\exp(-\epsilon_{{\bf k}0}^e \tau) 
-n_{{\bf k}0}^e\, \left[ \theta(-\tau) - N(\epsilon^e_{{\bf k}0}) \right] 
\exp(\epsilon_{{\bf k}0}^e \tau) \right\} ,\\
G_0^-(\tau,{\bf k}) & = & - \sum_{e = \pm} 
\gamma_0\, \Lambda^e_{\bf k} \left\{ \frac{}{}\! n_{{\bf k}0}^e\,
\left[ \theta(\tau) - N(\epsilon^e_{{\bf k}0}) \right] 
\exp(-\epsilon_{{\bf k}0}^e \tau) 
-(1-n_{{\bf k}0}^e)\, \left[ \theta(-\tau) - N(\epsilon^e_{{\bf k}0}) 
\right]  \exp(\epsilon_{{\bf k}0}^e \tau) \right\}\,\, .
\end{eqnarray}
\end{mathletters}
Here, $N(x) \equiv (e^{x/T} + 1)^{-1}$, and
\begin{equation}
n_{{\bf k}0}^e \equiv \frac{\epsilon_{{\bf k}0}^e + \mu - ek}{2 \,
\epsilon_{{\bf k}0}^e} 
\end{equation}
are the occupation numbers of particles ($e = +1$) or antiparticles
($e = -1$) at zero temperature. Consequently,
$1-n_{{\bf k}0}^e$ are the occupation numbers for particle-holes
or antiparticle-holes. 

Note that
\begin{equation} \label{chargeconjprop}
G_0^\pm(-\tau,{\bf k}) = - \gamma_0\, G_0^\mp(\tau,{\bf k}) \, \gamma_0 \,\,.
\end{equation}
For $0 \leq \tau \leq 1/T$, one derives with $1-N(x) = N(x) \,e^{x/T}$
\begin{equation} \label{KMS}
G_0^\pm \left(\frac{1}{T} - \tau, {\bf k}\right) 
= - G_0^\pm(-\tau,{\bf k}) \,\, ,
\end{equation}
the well-known Kubo--Martin--Schwinger relation for fermions \cite{LeBellac}.

Using the fact that $n_{{\bf k}0}^e \equiv \theta(\mu - ek)$, and
$N(x) = 1-N(-x)$, the propagators (\ref{mixedprops}) can be cast
into the more familiar form
\begin{mathletters}
\begin{eqnarray}
G_0^+(\tau,{\bf k}) & = & - \Lambda^+_{\bf k}\, \gamma_0 \, 
\left[ \theta(\tau) - N_F^+(k) \right]\, e^{-(k-\mu) \tau}
+\Lambda^-_{\bf k}\, \gamma_0 \, 
\left[ \theta(-\tau) - N_F^-(k) \right]\, e^{(k+\mu) \tau} \,\, , \\
G_0^-(\tau,{\bf k}) & = & \gamma_0\, \Lambda^+_{\bf k} \,
\left[ \theta(-\tau) - N_F^+(k) \right] \, e^{(k-\mu)\tau}
- \gamma_0\, \Lambda^-_{\bf k} \, 
\left[ \theta(\tau) - N_F^-(k) \right] \, e^{-(k+\mu)\tau}
\,\,,
\end{eqnarray}
\end{mathletters}
where $N_F^\pm(k) \equiv N(k\mp\mu)$ is the Fermi--Dirac distribution
function for particles (antiparticles).
However, in view of the application to the superconducting phase
in Sec.\ \ref{iv}, it is advantageous to continue to
use the form (\ref{mixedprops}).

Denoting $K_1 \equiv K$ and $K_2 \equiv K-P$,
one computes the expressions
\begin{equation} 
T \sum_{k_0} {\rm Tr}_s \left[ \gamma^\mu \, 
G_0^\pm(K_1) \, \gamma^\nu \, G_0^\pm(K_2) \right]
= T\sum_{k_0} \int_0^{1/T} d\tau_1 \, d\tau_2 \,
e^{k_0 \tau_1 + (k_0-p_0) \tau_2}\, 
{\rm Tr}_s \left[ \gamma^\mu \, G_0^\pm(\tau_1,{\bf k}_1) \,
\gamma^\nu \, G_0^\pm(\tau_2,{\bf k}_2) \right]
\label{expressions}
\end{equation}
as follows.
To perform the Matsubara sum over $k_0$, use the identity \cite{LeBellac}
\begin{equation} \label{identity}
T \sum_n e^{k_0 \tau} = \sum_{m = - \infty}^\infty (-1)^m \, 
\delta\left( \tau - \frac{m}{T} \right) \,\, ,
\end{equation}
valid for fermionic Matsubara frequencies, $k_0 = -i (2n+1) \pi T$.
Since $0 \leq \tau_1, \, \tau_2 \leq 1/T$ in Eq.\ (\ref{expressions}),
the delta function in Eq.\ (\ref{identity}) has support only for $m=1$,
{\it i.e.}, $\tau_2 = 1/T - \tau_1$. 
With the help of Eqs.\ (\ref{chargeconjprop}) and (\ref{KMS}), as well
as $e^{p_0/T} = 1$ for bosonic Matsubara frequencies $p_0 = -i 2n\pi T$, 
one obtains
\begin{equation}
T \sum_{k_0} {\rm Tr}_s \left[ \gamma^\mu \, G_0^\pm(K_1) \,
\gamma^\nu \, G_0^\pm(K_2) \right]
= - \int_0^{1/T} d \tau\, e^{p_0 \tau}\,
{\rm Tr}_s \left[ \gamma^\mu \, G_0^\pm(\tau,{\bf k}_1) \,
\gamma^\nu \, \gamma_0\, G_0^\mp(\tau,{\bf k}_2)\, \gamma_0 \right] \,\,.
\end{equation}
One now inserts the expressions (\ref{mixedprops}), and integrates 
over $\tau$. Putting everything together, one
obtains for the gluon self-energy:
\begin{eqnarray}
\lefteqn{ {\Pi_0}^{\mu \nu} (P)  =  - \frac{1}{4}\, g^2\, N_f
\, \int \frac{d^3{\bf k}}{(2 \pi)^3} \sum_{e_1,e_2 = \pm}  
 \left\{ \frac{}{}\!
{\cal T}_+^{\mu \nu}({\bf k}_1,{\bf k}_2) \right.} \nonumber \\
& \times &  \left[ \left( \frac{n^0_1 \, (1-n^0_2)}{p_0 + \epsilon^0_1
+ \epsilon^0_2} - \frac{(1-n^0_1)\, n^0_2}{p_0 - \epsilon^0_1
- \epsilon^0_2} \right)\! (1-N^0_1 - N^0_2) +
 \left( \frac{(1-n^0_1)\, (1-n^0_2)}{p_0 - \epsilon^0_1
+ \epsilon^0_2} - \frac{n^0_1\, n^0_2}{p_0 + \epsilon^0_1
- \epsilon^0_2} \right)\! (N^0_1 - N^0_2) \right] \nonumber \\
&   & \hspace*{5.2cm} +  \; {\cal T}_-^{\mu \nu}({\bf k}_1,{\bf k}_2)
\nonumber \\
& \times &  \left. \!
\left[ \left( \frac{(1-n^0_1)\, n^0_2}{p_0 + \epsilon^0_1
+ \epsilon^0_2} - \frac{n^0_1\, (1- n^0_2)}{p_0 - \epsilon^0_1
- \epsilon^0_2} \right) \! (1-N^0_1 - N^0_2) 
+  \left( \frac{n^0_1 \, n^0_2}{p_0 - \epsilon^0_1
+ \epsilon^0_2} - \frac{(1-n^0_1)\,(1- n^0_2)}{p_0 + \epsilon^0_1
- \epsilon^0_2} \right) \! (N^0_1 - N^0_2) \right]  \right\}  . \!
\label{Pi} 
\end{eqnarray}
Here, 
\begin{equation} \label{traces}
{\cal T}_\pm^{\mu \nu}({\bf k}_1,{\bf k}_2) \equiv
{\rm Tr}_s \left( \gamma_0 \, \gamma^\mu \, 
\Lambda^{\pm e_1}_{{\bf k}_1} \, \gamma_0\, \gamma^\nu \, 
\Lambda^{\pm e_2}_{{\bf k}_2} \right)\,\, ,
\end{equation}
and I introduced the somewhat compact notation
\begin{equation}
\epsilon^0_i \equiv \epsilon_{{\bf k}_i0}^{e_i} \;\;\; , \;\;\;\;
n^0_i \equiv n_{{\bf k}_i0}^{e_i}\;\;\;, \;\;\;\;
N^0_i \equiv N(\epsilon^0_i) \,\, .
\end{equation}
An (appropriately generalized)
expression of the form (\ref{Pi}) will also appear in 
Sec.\ \ref{iv}, when the self-energy is computed in the superconducting
phase. In the normal phase, however, one can use 
$n_i^0 \equiv \theta(\mu - e_i k_i)$ to show that
\begin{mathletters}
\begin{eqnarray}
n_i^0\, (1-N_i^0) & = & n_i^0 \, 
         \left\{ \theta(e_i)\, N_F^+(k_i)
               + \theta(-e_i)\, \left[ 1-N_F^-(k_i) \right] \right\}\,\, ,\\
(1-n_i^0)\, N_i^0 & = & (1-n_i^0)\, 
          \left\{ \theta(e_i)\, N_F^+(k_i) 
                + \theta(-e_i)\, \left[ 1-N_F^-(k_i) \right] \right\}\,\, \\
(1-n_i^0)\, (1-N_i^0) & = & (1- n_i^0) \, 
          \left\{ \theta(e_i)\, \left[1-N_F^+(k_i) \right] 
                + \theta(-e_i)\, N_F^-(k_i) \right\} \,\, ,\\
n_i^0\, N_i^0 & = & n_i^0 \, 
         \left\{ \theta(e_i)\, \left[1-N_F^+(k_i) \right] 
               + \theta(-e_i)\, N_F^-(k_i) \right\}\,\, .
\end{eqnarray}
\end{mathletters}
Equation (\ref{Pi}) then simplifies to
\begin{eqnarray}
{\Pi_0}^{\mu \nu} (P) & = & \frac{1}{4}\,  g^2\, N_f
\, \int \frac{d^3{\bf k}}{(2 \pi)^3} \sum_{e_1,e_2 = \pm}  
\left[ \frac{{\cal T}_+^{\mu \nu}({\bf k}_1,{\bf k}_2)}{
p_0 - e_1 \, k_1 + e_2\,  k_2}
-  \frac{{\cal T}_-^{\mu \nu}({\bf k}_1, {\bf k}_2)}{
p_0 + e_1 \, k_1 - e_2\, k_2} \right]
\nonumber \\
&   & \times \, \left\{ \frac{}{}
       \theta(e_1) \left[ 1-N_F^+(k_1) \right] + \theta(-e_1)\, N_F^-(k_1)
      -\theta(e_2) \left[ 1-N_F^+(k_2) \right] - \theta(-e_2)\, N_F^-(k_2)
 \right\} \,\, .
\label{Pi2}
\end{eqnarray}

\subsection{Trace over spinor space} \label{iiie}

The traces (\ref{traces}) are best computed for temporal and spatial
components separately,
\begin{mathletters} \label{traces2}
\begin{eqnarray}
{\cal T}_\pm^{00} & = & 1 + e_1 e_2\, \hat{\bf k}_1 \cdot \hat{\bf k}_2
\,\, , \\
{\cal T}_\pm^{0i}  =  {\cal T}_\pm^{i0} 
& = & \pm e_1 \, \hat{k}_1^i \pm e_2 \, \hat{k}_2^i 
\;\;\; , \;\;\;\; i = x,y,z \,\, , \\
{\cal T}_\pm^{ij} & = & \delta^{ij} \, 
\left(1- e_1 e_2\,\hat{\bf k}_1 \cdot \hat{\bf k}_2\right) + e_1 e_2 \,
\left(\hat{k}_1^i \, \hat{k}_2^j + \hat{k}_1^j \, \hat{k}_2^i\right)
\;\;\; , \;\;\;\; i,j = x,y,z \,\, .
\end{eqnarray}
\end{mathletters}
Equation (\ref{Pi}), or Eq.\ (\ref{Pi2}), together with Eqs.\ (\ref{traces2}),
completes the computation of the quark contribution to the
gluon self-energy to one-loop order in the normal phase.
At temperatures $T \ll \mu$, this is the dominant contribution
to the gluon self-energy. In the following, I study the so-called
hard-dense-loop (HDL) limit.

\subsection{The HDL limit} \label{HDL}

To derive the HDL limit, it is advantageous to shift
the integration over 3-momentum in Eqs.\ (\ref{Pi}) or (\ref{Pi2}),
${\bf k} \rightarrow {\bf k} + {\bf p}/2$, such that
${\bf k}_1 = {\bf k}+{\bf p}/2$ and ${\bf k}_2 = {\bf k}-{\bf p}/2$.
The HDL limit is obtained by taking $p_0, p$ to be of order
$g \mu$ (``soft''), while $k$ is of order
$\mu$ (``hard'') \cite{LeBellac}.
As the gluon self-energy (\ref{Pi}) is already proportional to $g^2$, it is 
permissible to compute the integral in Eq.\ (\ref{Pi}) 
to order $O(p^0)$. However, since some of the energy denominators 
are of order $O(p)$,
one has to keep terms up to order $O(p)$ in the numerators, too.
For the traces (\ref{traces2}) one then obtains
\begin{mathletters}
\begin{eqnarray}
{\cal T}_\pm^{00} & \simeq & 1 + e_1 e_2 + 
O\left(\frac{p^2}{k^2} \right) \,\, , \label{43a} \\
{\cal T}_\pm^{0i}  =  {\cal T}_\pm^{i0} 
& \simeq & 
\pm (e_1 + e_2) \, \hat{k}^i  \pm (e_1 - e_2)\, \left(\delta^{ij} -
\hat{k}^i\, \hat{k}^j\right) \frac{p^j}{2\,k} + 
O\left(\frac{p^2}{k^2} \right)\,\, , \label{43b} \\
{\cal T}_\pm^{ij} & \simeq & \delta^{ij} \, (1- e_1 e_2) + 2\, e_1 e_2 \,
\hat{k}^i \, \hat{k}^j + O\left(\frac{p^2}{k^2} \right)\,\, .
\label{43c}
\end{eqnarray}
\end{mathletters}
In the following, the temporal and spatial components of the
gluon self-energy are evaluated separately.
\\ ~~ \\
\noindent
\underline{(i) $\mu = \nu =0$:}\hspace*{0.5cm}
In the HDL limit, Eq.\ (\ref{43a}) shows that only
particle-particle ($e_1 = e_2 = +1)$, or antiparticle-antiparticle 
($e_1 = e_2 =-1$) excitations contribute
to the electric components of the gluon self-energy.
In this case, only the difference $k_1 - k_2$ occurs in
the energy denominators in Eq.\ (\ref{Pi2}), which, in the HDL limit, is 
\begin{equation} \label{k1-k2}
k_1 - k_2 \simeq {\bf p} \cdot \hat{\bf k} \,\, .
\end{equation}
In the numerators, the difference of the thermal occupation numbers is
\begin{equation} \label{N1-N2}
N_F^\pm(k_1) - N_F^\pm(k_2) \simeq {\bf p} \cdot \hat{\bf k}
\; \frac{d N_F^\pm(k)}{dk} \,\, .
\end{equation}
Equation (\ref{Pi2}) with Eq.\ (\ref{43a}) then yields
\begin{equation}
{\Pi_0}^{00} (P) \simeq   g^2\, N_f
\, \int \frac{d^3{\bf k}}{(2 \pi)^3} 
\left( 1- \frac{p_0}{p_0 + {\bf p} \cdot \hat{\bf k}} \right)
\, \left[ \frac{d N_F^+(k)}{dk} + \frac{d N_F^-(k)}{dk} \right] \,\, .
\label{Pi3}
\end{equation}
With some effort, one can also do the integration over $k$ 
exactly for {\em all\/} temperatures and
chemical potentials \cite{LeBellac}. In this case,
the final answer encompasses not only the hard-dense-loop
limit, but also the ``hard-thermal-loop'' (HTL) limit.
That much effort is, however, 
not necessary in the present case. For superconductivity,
one is interested in temperatures of the order 
of the zero-temperature gap, $T \sim \phi_0 \sim \mu \,
\exp(-c_{\rm QCD}/g) \ll \mu$. On this basis
it was argued above that contributions from the gluon and ghost loops 
to the gluon self-energy can be neglected, as they are $\sim g^2 T^2$,
while the dominant contribution from the quark loop is $\sim g^2 \mu^2$.

In essence this means that effects from nonzero temperature can be
neglected to leading order. Consequently,
\begin{equation} \label{Fermisurface}
\frac{d N_F^+(k)}{dk} \simeq \frac{d \,\theta(\mu -k) }{dk} = - \delta(k- \mu)
\;\;\; , \;\;\;\; 
\frac{d N_F^-(k)}{dk} \simeq \frac{d \,\theta(k+\mu) }{dk} = 0\,\, .
\end{equation}
From the physical point of view this is an important relation:
only quark excitations {\em at\/} the Fermi surface contribute to
the gluon self-energy.

With these approximations one obtains the well-known result \cite{LeBellac}
\begin{equation}
{\Pi_0}^{00} (P) \simeq -  3\, m_g^2
\int \frac{d\Omega}{4 \pi} 
\left( 1- \frac{p_0}{p_0 + {\bf p} \cdot \hat{\bf k}} \right) \,\, ,
\label{Pi4}
\end{equation}
where $d\Omega$ is the integration over solid angle and
\begin{equation} \label{gluonmass}
m_g^2 \equiv g^2\, \frac{N_f}{6 \pi^2}  \, \mu^2
\end{equation}
is the gluon mass at $T=0$.
Equation (\ref{Pi4}) remains valid in the HTL limit, when Eq.\ 
(\ref{gluonmass}) is properly generalized to nonzero
temperature \cite{LeBellac}.

In the static limit, $p_0 = 0$, the dependence on ${\bf p}$ vanishes,
and one simply has
\begin{equation} \label{Pi00HDLstatic}
{\Pi_0}^{00} (0) \simeq - 3\, m_g^2\,\, ,
\end{equation}
the usual result for Debye screening.
\\ ~~ \\
\noindent
\underline{(ii) $\mu = 0,\, \nu =i$:}\hspace*{0.5cm} 
For ${\Pi_0}^{0i}$, one concludes from Eqs.\ (\ref{Pi2}) and
(\ref{43b}) that particle-antiparticle excitations are at
least of order $O(p^2)$, {\it i.e.}, to leading order in the HDL limit
only particle-particle or antiparticle-antiparticle
excitations contribute to the gluon self-energy.
Then, with Eqs.\ (\ref{k1-k2}) and (\ref{N1-N2}) one obtains
\begin{equation}
{\Pi_0}^{0i} (P) \simeq   g^2\, N_f
\, \int \frac{d^3{\bf k}}{(2 \pi)^3} 
\,\frac{p_0\, \hat{k}^i }{p_0 + {\bf p} \cdot \hat{\bf k}}
\, \left[ \frac{d N_F^+(k)}{dk} + \frac{d N_F^-(k)}{dk} \right] \,\, .
\label{Pi5}
\end{equation}
For the temperatures of interest, one can again
make the approximation (\ref{Fermisurface}) to obtain
\begin{equation}
{\Pi_0}^{0i} (P) \simeq  -  3\, m_g^2
\, \int \frac{d\Omega}{4 \pi} 
\,\frac{p_0\, \hat{k}^i }{p_0 + {\bf p} \cdot \hat{\bf k}} 
\,\, ,
\label{Pi6}
\end{equation}
which coincides with \cite{LeBellac}.

In the static limit,
\begin{equation} \label{Pi0iHDLstatic}
{\Pi_0}^{0i} (0) \simeq  0\,\, .
\end{equation}
\\ ~~ \\
\noindent
\underline{(iii) $\mu = i,\, \nu =j$:}\hspace*{0.5cm} 
For ${\Pi_0}^{ij}$, Eq.\ (\ref{43c}) shows that
not only particle-particle ($e_1 = e_2 = +1$) 
and antiparticle-antiparticle ($e_1 = e_2 = -1$) excitations
contribute, as for the other components of ${\Pi_0}^{\mu \nu}$, but
also particle-antiparticle ($e_1 = -e_2 = \pm 1$) excitations.
In the former, one encounters again the difference of momenta (\ref{k1-k2})
and thermal occupation numbers (\ref{N1-N2}). In the latter, however,
the sum of momenta and thermal occupation numbers occurs. To leading
order in $p$, 
\begin{equation}
k_1 + k_2 \simeq 2\, k\;\;\; , \;\;\;\;
N_F^\pm(k_1) + N_F^\pm(k_2) \simeq  2\, N_F^\pm(k)\,\,.
\end{equation}
Then,
\begin{eqnarray}
{\Pi_0}^{ij}(P) & \simeq &  g^2\, N_f
\, \int \frac{d^3{\bf k}}{(2 \pi)^3} \left\{
\hat{k}^i \, \hat{k}^j \,
\left( 1- \frac{p_0}{p_0 + {\bf p} \cdot \hat{\bf k}} \right)
\, \left[ \frac{d N_F^+(k)}{dk} + \frac{d N_F^-(k)}{dk} \right]
\right. \nonumber \\
&    & \left. \hspace*{2.1cm}
- \left(\delta^{ij} - \hat{k}^i \, \hat{k}^j \right)\,
\frac{1}{k} \left[1 - N_F^+(k) - N_F^-(k) \right] \right\} \,\, .
\label{Piij}
\end{eqnarray}
The 1 in the last term is an ultraviolet-divergent vacuum contribution 
and has to be removed by renormalization. 
The angular integration can be performed for the parts which do not depend
on ${\bf p}$, $\int (d\Omega/ 4\pi)\, \hat{k^i}\, \hat{k^j}
\equiv \delta^{ij}/3$. One then realizes after an integration by parts
that the ${\bf p}$-independent part of the first line in Eq.\ (\ref{Piij})
cancels the second line,
\begin{equation}
{\Pi_0}^{ij}(P)  \simeq  -  g^2\, N_f
\, \int \frac{d^3{\bf k}}{(2 \pi)^3} \, \hat{k}^i \, \hat{k}^j \,
\frac{p_0}{p_0 + {\bf p} \cdot \hat{\bf k}} 
\, \left[ \frac{d N_F^+(k)}{dk} + \frac{d N_F^-(k)}{dk} \right]\,\,.
\end{equation}
With the gluon mass (\ref{gluonmass}) this can be written in the form
\begin{equation}
{\Pi_0}^{ij}(P)  \simeq  3 \, m_g^2
\, \int \frac{d\Omega}{4 \pi} \, \hat{k}^i \, \hat{k}^j \,
\frac{p_0}{p_0 + {\bf p} \cdot \hat{\bf k}} \,\, ,
\end{equation}
which is the standard result \cite{LeBellac}.
Static magnetic gluons are not screened,
\begin{equation} \label{PiijHDLstatic}
{\Pi_0}^{ij}(0)  \simeq  0 \,\, .
\end{equation}

\section{The gluon self-energy in the superconducting phase}\label{iv}

In the superconducting phase, the ground state is a condensate
of quark Cooper pairs, $\langle \bar{\psi}_C \, \psi \rangle \neq 0$.
As was shown in \cite{prscalar}, in mean-field approximation the
quark propagator (\ref{S0K}) becomes
\begin{equation} \label{SK}
{\cal S}(K) = \left( \begin{array}{cc}
             G^+(K) & \Xi^-(K) \\
             \Xi^+(K) & G^-(K) 
       \end{array} \right)\,\, ,
\end{equation}
where the quasiparticle and charge conjugate quasiparticle propagators
are
\begin{equation}
G^\pm \equiv \left( G_0^\pm - \Sigma^\pm \right)^{-1}\;\;\; , \;\;\;\;
\Sigma^\pm \equiv \Phi^\mp \, G_0^\mp \, \Phi^\pm \,\, .
\end{equation}
$\Sigma^\pm$ is the quark self-energy generated by exchanging particles
or charge conjugate particles with the condensate. 
For $\Sigma^+$, a particle annihilates with an antiparticle 
in the condensate $\Phi^+ \sim \langle \psi_C \, \bar{\psi} \rangle$, 
and a charge conjugate particle is created.
This charge conjugate particle propagates via $G_0^-$, until it 
annihilates in the condensate $\Phi^- \sim \langle \psi \, \bar{\psi}_C
\rangle$ with a charge conjugate antiparticle, 
whereby a particle is created \cite{prqcd}. 
The meaning of $\Sigma^-$ can be explained analogously,
except that the roles of particles and charge conjugate particles
are interchanged.

The off-diagonal components of the quark propagator (\ref{SK}) are
\begin{equation} \label{offdiagonal}
\Xi^\pm \equiv - G_0^\mp \, \Phi^\pm \, G^\pm \,\, .
\end{equation}
The physical interpretation is that particles (charge
conjugate particles) annihilate with an antiparticle (a charge
conjugate antiparticle) in the condensate, upon which a
charge conjugate particle (a particle) is created.

In mean-field approximation, the condensate $\Phi^+$ obeys the gap equation
\cite{prlett2,prqcd,prscalar}
\begin{equation} \label{gapequation}
\Phi^+(K) \equiv - g^2 \, \frac{T}{V} \sum_Q \Delta^{ab}_{\mu \nu} (K-Q)\,
\bar{\Gamma}^\mu_a \, \Xi^+(Q) \, \Gamma^\nu_b \,\, ,
\end{equation}
and $\Phi^-$ can be obtained from
\begin{equation} \label{phi-phi+}
\Phi^-(K) \equiv \gamma_0 \left[ \Phi^+(K) \right]^\dagger \gamma_0\,\, .
\end{equation}
The solution of the gap equation (\ref{gapequation}) has been
extensively discussed in \cite{prqcd}.

The gluon self-energy (\ref{PiP}) becomes
\begin{equation} \label{PiPscfNG}
\Pi^{\mu \nu}_{ab} (P) = \frac{1}{2} \,g^2 \, \frac{T}{V} 
\sum_K {\rm Tr}_{s,c,f,NG} \left[ \hat{\Gamma}^\mu_a
\, {\cal S} (K) \, \hat{\Gamma}^\nu_b \, {\cal S}(K-P) \right]
\,\,.
\end{equation}
As in the normal phase, this expression is computed in several steps.

\subsection{Trace over Nambu--Gor'kov space}

The trace over the 2-dimensional Nambu--Gor'kov space is readily
performed with Eqs.\ (\ref{Gamma}) and (\ref{SK}),
\begin{eqnarray}
\Pi^{\mu \nu}_{ab} (P) & = & \frac{1}{2} \,g^2 \, \frac{T}{V} 
\sum_K {\rm Tr}_{s,c,f} \left[ \Gamma^\mu_a \, G^+ (K) \, \Gamma^\nu_b 
\, G^+(K-P) + \bar{\Gamma}^\mu_a \, G^- (K) \, \bar{\Gamma}^\nu_b 
\, G^-(K-P) \right. \nonumber \\
&   & \left. \hspace*{2.55cm}
+ \Gamma^\mu_a \, \Xi^- (K) \, \bar{\Gamma}^\nu_b 
\, \Xi^+(K-P) + \bar{\Gamma}^\mu_a \, \Xi^+ (K) \, \Gamma^\nu_b 
\, \Xi^-(K-P)  \right] \,\,. \label{PiPscf}
\end{eqnarray}
When the temperature approaches the critical temperature, 
$T \rightarrow T_c$, the condensate melts, $\Phi^\pm \rightarrow 0$, 
{\it i.e.}, $\Xi^\pm \rightarrow 0$ and $G^\pm \rightarrow G_0^\pm$,
and the gluon self-energy assumes the
form of the normal phase, 
$\Pi^{\mu \nu}_{ab} \rightarrow {\Pi_0}^{\mu \nu}_{ab}$,
which was discussed in the previous Sec.\ \ref{iii}.

\subsection{Trace over flavor space}

For a condensate with total spin $J=0$ and
$N_f =2$, the condensate is totally antisymmetric in
flavor space \cite{prlett},
\begin{equation}
\left( \Phi^\pm \right)_{fg} \equiv \pm \epsilon_{fg} \, \Phi^\pm \,\, ,
\end{equation}
where use has been made of 
$\epsilon_{fg}^T = \epsilon_{gf} = - \epsilon_{fg}$.
Consequently, since the free quark propagator is diagonal in flavor space,
the quark self-energy is also diagonal in flavor space,
\begin{equation}
\left( \Sigma^\pm \right)_{fg} = \left( \Phi^\mp \right)_{fh} \,
\left( G_0^\mp \right)_{hm} \, \left( \Phi^\pm \right)_{mg} 
= \epsilon_{hf} \, \epsilon_{hg} \, \Phi^\mp \, G_0^\mp \, \Phi^\pm
= \delta_{fg}\, \Sigma^\pm \,\,.
\end{equation}
Then, also the quasiparticle propagator is diagonal in flavor space,
\begin{equation}
\left( G^\pm \right)_{fg} = \delta_{fg}\, G^\pm \,\, .
\end{equation}
On the other hand, the off-diagonal components of ${\cal S}$ are
antisymmetric in flavor space,
\begin{equation}
\left( \Xi^\pm \right)_{fg} = - \left( G_0^\mp \right)_{fh} \, 
\left( \Phi^\pm \right)_{hm} \, \left( G^\pm  \right)_{mg} 
= \pm \epsilon_{fg} \, \Xi^\pm \,\, .
\end{equation}
As the vertices $\Gamma^\mu_a$ and $\bar{\Gamma}^\mu_a$ are
flavor-blind (proportional to the unit matrix in flavor space),
the trace over flavor space in Eq.\ (\ref{PiPscf}) results in
\begin{eqnarray}
\Pi^{\mu \nu}_{ab} (P) & = & \frac{1}{2} \,g^2 \,N_f\, \frac{T}{V} 
\sum_K {\rm Tr}_{s,c} \left[ \Gamma^\mu_a \, G^+ (K) \, \Gamma^\nu_b 
\, G^+(K-P) + \bar{\Gamma}^\mu_a \, G^- (K) \, \bar{\Gamma}^\nu_b 
\, G^-(K-P) \right. \nonumber \\
&   & \left. \hspace*{2.85cm}
+ \Gamma^\mu_a \, \Xi^- (K) \, \bar{\Gamma}^\nu_b 
\, \Xi^+(K-P) + \bar{\Gamma}^\mu_a \, \Xi^+ (K) \, \Gamma^\nu_b 
\, \Xi^-(K-P)  \right] \,\,, \label{PiPsc}
\end{eqnarray}
where, of course, $N_f = 2$.

\subsection{Trace over color space}

A $N_f =2$, $J=0$ condensate is also totally antisymmetric in
color space \cite{prlett},
\begin{equation}
\left( \Phi^\pm \right)_{ij} \equiv \pm \epsilon_{ij3} \, \Phi^\pm \,\, ,
\end{equation}
where use has been made of $\epsilon_{ij3}^T = \epsilon_{ji3} = 
- \epsilon_{ij3}$, and where a
global color rotation has been performed to orient the condensate
into the (anti-)3--direction in color space.
(The notation is again sloppy: the ``3'' is actually not a
triplet, but an anti-triplet index.)

The free quark propagator is diagonal in color space,
so that one computes for the quark self-energy:
\begin{equation}
\left( \Sigma^\pm \right)_{ij} = \left( \Phi^\mp \right)_{ik} \,
\left( G_0^\mp \right)_{kl} \, \left( \Phi^\pm \right)_{lj} 
= \epsilon_{ki3} \, \epsilon_{kj3} \, \Phi^\mp \, G_0^\mp \, \Phi^\pm
= \left( \delta_{ij} - \delta_{i3}\, \delta_{j3} \right)
\, \Sigma^\pm \,\,.
\end{equation}
This result is physically easy to interpret, remembering the above
discussion of how the quark self-energy arises. Quarks with
color 3 do not condense, consequently there is no antiquark in
the condensate which a  color--3 quark could annihilate with. Thus, 
color--3 quarks do not attain a self-energy \cite{prqcd}.

The color structure of the quasiparticle propagator is therefore
\begin{equation}
\left( G^\pm \right)_{ij} = 
\left( \delta_{ij} - \delta_{i3} \, \delta_{j3} \right)\,
G^\pm + \delta_{i3} \, \delta_{j3} \, G_0^\pm \,\, .
\end{equation}
For the off-diagonal components of ${\cal S}$ one then finds
\begin{equation}
\left( \Xi^\pm \right)_{ij} = - \left( G_0^\mp \right)_{ik} \, 
\left( \Phi^\pm \right)_{kl} \, \left( G^\pm  \right)_{lj} 
= \pm \epsilon_{ij3} \, \Xi^\pm \,\, .
\end{equation}
One now computes the trace over color space with the explicit form of the 
Gell-Mann matrices. After a somewhat tedious, but straightforward
calculation one obtains for $a=b=1,2,3$:
\begin{mathletters} \label{Pisupercond}
\begin{eqnarray}
\Pi^{\mu \nu}_{11} (P) & = & \frac{1}{4} \,g^2 \,N_f\, \frac{T}{V} 
\sum_K {\rm Tr}_{s} \left[ \gamma^\mu \, G^+ (K) \, \gamma^\nu 
\, G^+(K-P) + \gamma^\mu \, G^- (K) \, \gamma^\nu 
\, G^-(K-P) \right. \nonumber \\
&   & \left. \hspace*{2.65cm}
+ \gamma^\mu \, \Xi^- (K) \, \gamma^\nu
\, \Xi^+(K-P) + \gamma^\mu \, \Xi^+ (K) \, \gamma^\nu 
\, \Xi^-(K-P)  \right] \,\,, \label{Pi11}
\end{eqnarray}
for $a=b=4,5,6,7$:
\begin{eqnarray}
\Pi^{\mu \nu}_{44} (P) & = & \frac{1}{8} \,g^2 \,N_f\, \frac{T}{V} 
\sum_K {\rm Tr}_{s} \left[ \gamma^\mu \, G_0^+ (K) \, \gamma^\nu 
\, G^+(K-P) + \gamma^\mu \, G^+ (K) \, \gamma^\nu 
\, G_0^+(K-P) \right. \nonumber \\
&   & \left. \hspace*{2.65cm}
+ \gamma^\mu \, G_0^- (K) \, \gamma^\nu
\, G^-(K-P) + \gamma^\mu \, G^- (K) \, \gamma^\nu 
\, G_0^-(K-P)  \right] \,\,, \label{Pi44}
\end{eqnarray}
and for $a=b=8$:
\begin{eqnarray}
\Pi^{\mu \nu}_{88} (P) & = & \frac{2}{3} \, {\Pi_0}^{\mu \nu}(P) 
+ \frac{1}{3} \, \tilde{\Pi}^{\mu \nu} (P) \,\, , \nonumber \\
\tilde{\Pi}^{\mu \nu} (P) & = & \frac{1}{4} \,g^2 \,N_f\, \frac{T}{V} 
\sum_K {\rm Tr}_{s} \left[ \gamma^\mu \, G^+ (K) \, \gamma^\nu 
\, G^+(K-P) + \gamma^\mu \, G^- (K) \, \gamma^\nu 
\, G^-(K-P) \right. \nonumber \\
&   & \left. \hspace*{2.65cm}
- \gamma^\mu \, \Xi^- (K) \, \gamma^\nu
\, \Xi^+(K-P) - \gamma^\mu \, \Xi^+ (K) \, \gamma^\nu 
\, \Xi^-(K-P)  \right] \,\,, \label{Pi88}
\end{eqnarray}
where ${\Pi_0}^{\mu \nu}$ is the gluon self-energy in the normal phase,
Eq.\ (\ref{Pisb}).

Apart from the diagonal elements (\ref{Pi11}) -- (\ref{Pi88}), after
performing the color-trace one also finds the off-diagonal elements
\begin{eqnarray}
\Pi^{\mu \nu}_{45} (P) & = & - \Pi^{\mu \nu}_{54} (P) =
\Pi^{\mu \nu}_{67} (P) = -  \Pi^{\mu \nu}_{76}(P) \equiv i \,
\hat{\Pi}^{\mu \nu} (P) \,\, ,\nonumber \\
\hat{\Pi}^{\mu \nu}(P)& \equiv &  \frac{1}{8} \,g^2 \,N_f\, \frac{T}{V} 
\sum_K {\rm Tr}_{s} \left[ \gamma^\mu \, G_0^+ (K) \, \gamma^\nu 
\, G^+(K-P) - \gamma^\mu \, G^+ (K) \, \gamma^\nu 
\, G_0^+(K-P) \right. \nonumber \\
&   & \left. \hspace*{2.6cm}
- \gamma^\mu \, G_0^- (K) \, \gamma^\nu
\, G^-(K-P) + \gamma^\mu \, G^- (K) \, \gamma^\nu 
\, G_0^-(K-P)  \right] \,\,. \label{Pi45}
\end{eqnarray}
\end{mathletters}
The occurrence of these off-diagonal elements bears no special physical
meaning. It simply indicates that the inverse gluon propagator 
$\Delta^{-1}$ is not diagonal in the original basis of adjoint colors.
For instance, in the $(45)$-subspace of adjoint colors
$\Delta^{-1}$ has the form
\begin{equation}
 \left( \begin{array}{cc}
\Delta_0^{-1} + \Pi_{44} & i \, \hat{\Pi} \\
-i\, \hat{\Pi} & \Delta_0^{-1} + \Pi_{44} 
\end{array} \right) \,\, .
\end{equation}
This hermitean matrix is easily diagonalized by the unitary matrix
\begin{equation} \label{Utrafo}
U \equiv \frac{1}{\sqrt{2}} \left( \begin{array}{cc}
           1 & -i \\
           -i & 1 
\end{array} \right) \,\, .
\end{equation}
In the new (diagonal) basis of adjoint colors,
\begin{equation}
 \left( \begin{array}{cc}
\Delta_0^{-1} + \Pi_{44} + \hat{\Pi} & 0 \\
0 & \Delta_0^{-1} + \Pi_{44} - \hat{\Pi} 
\end{array} \right) \,\, .
\end{equation}
Similar arguments hold for the $(67)$-subspace.
Therefore, rotating into the new (diagonal) basis, 
\begin{equation} 
\Pi_{44} + \hat{\Pi} = \Pi_{66} + \hat{\Pi} \rightarrow \Pi_{44} = \Pi_{66}
\;\;\; , \;\;\;\;
\Pi_{44} - \hat{\Pi} = \Pi_{66} - \hat{\Pi} \rightarrow \Pi_{55} = \Pi_{77}
\,\, .
\end{equation}
In the following, only these diagonal gluon self energies will
be considered. They read explicitly
\begin{mathletters} \label{Pidiag}
\begin{eqnarray}
\Pi^{\mu \nu}_{44} (P) & = & \frac{1}{4} \,g^2 \,N_f\, \frac{T}{V} 
\sum_K {\rm Tr}_{s} \left[ \gamma^\mu \, G_0^+ (K) \, \gamma^\nu 
\, G^+(K-P) + \gamma^\mu \, G^- (K) \, \gamma^\nu 
\, G_0^-(K-P) \right]\,\, , \label{Pi44diag} \\
\Pi^{\mu \nu}_{55} (P) & = & \frac{1}{4} \,g^2 \,N_f\, \frac{T}{V} 
\sum_K {\rm Tr}_{s} \left[ \gamma^\mu \, G^+ (K) \, \gamma^\nu 
\, G_0^+(K-P) + \gamma^\mu \, G_0^- (K) \, \gamma^\nu 
\, G^-(K-P) \right]\,\, . \label{Pi55diag}
\end{eqnarray}
\end{mathletters}

Remembering the explicit form of the Gell-Mann matrices, the results 
(\ref{Pi11}), (\ref{Pi88}), and (\ref{Pidiag})
are simple to interpret. Gluons of adjoint colors 1, 2, and 3
see only quarks in the condensate, with fundamental colors 1 and 2. Their
self-energy has therefore contributions from the diagonal ($G^\pm$), 
as well as the off-diagonal ($\Xi^\pm$) components of the quark 
propagator (\ref{SK}).

Gluons of colors 4 and 5 ``see'' the uncondensed quark with fundamental
color 3, but also the condensed quarks of color 1. 
Analogously, gluons of colors 6 and 7 see the uncondensed quark and
the condensed quark of color 2. Therefore, the fermion
loop in the self-energy contains one free propagator $G_0^\pm$, corresponding
to the uncondensed quark, and one quasiparticle (charge conjugate
quasiparticle) propagator $G^\pm$, corresponding to the quark in the 
condensate. As there is no way to annihilate a color-3 quark in the
condensate, there is no contribution from the off-diagonal components
of Eq.\ (\ref{SK}).

Finally, gluons of color 8 see the condensed quarks of
colors 1 and 2, but also the uncondensed color-3 quark. 
The contribution to the gluon self-energy from the latter is equal to that in 
the normal phase, ${\Pi_0}^{\mu \nu}$,
the factor $2/3$ comes from the $(33)$-element of $T^8$.
Apart from the prefactor $1/3$, the contribution from the 
condensed quarks, $\tilde{\Pi}^{\mu \nu}$,
looks similar to $\Pi^{\mu \nu}_{11}$, except that the
sign of the last two terms is different. As will be seen below,
this difference is important to keep gluons of colors 1, 2, and 3 massless,
while the eighth gluon becomes massive.
Note that, for QED, $\bar{\Gamma}^\mu_a \rightarrow
\bar{\Gamma}^\mu = - \gamma^\mu$, $g \rightarrow e$. Thus,
for $N_f =2$,
the contribution from the condensed quarks to the self-energy of gluons
of color 8, $\tilde{\Pi}^{\mu \nu}$, is exactly $g^2/e^2$ of what 
one expects for the photon self-energy in an ordinary superconductor.

\subsection{Mixed representations for the quark propagators}

For $m=0$, the quasiparticle propagator can be written in
terms of chirality and energy projectors \cite{prqcd,prscalar},
\begin{equation} \label{quasiparticleprop}
G^\pm (K) = \sum_{h = r, \ell} \sum_{e = \pm} \frac{{\cal P}_{\pm h}\,
\Lambda^{\pm e}_{\bf k}}{k_0^2 - [\epsilon^e_{\bf k}(\phi_h^e)]^2} \,
\left[ G_0^\mp(K) \right]^{-1} \,\, ,
\end{equation}
where ${\cal P}_{r,\ell} = (1 \pm \gamma_5)/2$ are chirality
projectors (the notation $- h$ stands for $\ell$, if $h = r$, and
$r$, if $h = \ell$).
The quasiparticle energies are
\begin{equation} \label{quasiparticleenergy}
\epsilon_{\bf k}^e ( \phi_h^e ) \equiv \sqrt{(\mu - ek)^2 + | \phi_h^e |^2}
\,\, ,
\end{equation}
where $\phi_h^e$ is the gap function for pairing of quarks ($e = +1$)
or antiquarks ($e= -1$) with chirality $h$.

An analysis of the gap functions in mean-field approximation 
shows \cite{prqcd} that left- and right-handed gap functions differ
only by a complex phase factor,
\begin{equation} \label{phases}
\phi_r^e = \phi^e \, \exp(i\, \theta^e ) \;\;\; , \;\;\;\;
\phi_\ell^e = - \phi^e \, \exp(-i\, \theta^e )\,\, ,
\end{equation}
with $\phi^e \in {\bf R}$.
Moreover, the phase factor is independent of the energy projection,
$\theta^+ = \theta^- \equiv \theta$. Condensation fixes the value
of $\theta$, and breaks the $U_A(1)$ symmetry (which is effectively
restored at high densities) spontaneously.
If $\theta=0$ or $\pi/2$, condensation occurs in a spin-zero channel with good
parity, $J^P= 0^+$ or $J^P=0^-$, respectively. 
For $\theta \neq 0$, there is always a $J^P=0^-$ admixture, thus
condensation breaks also parity \cite{prlett,parity}. 
For the sake of simplicity, in the following we only consider 
$\theta =0$.

From Eq.\ (\ref{phases}), 
$|\phi_r^e| \equiv |\phi_\ell^e|  \equiv \phi^e$, and the sum over
chiralities in Eq.\ (\ref{quasiparticleprop}) is superfluous.
Writing the inverse free propagator as
$\left[ G_0^\mp (K) \right]^{-1} = \left[ k_0 \mp (\mu -ek) \mp
2\, ek \, \Lambda^{\mp e}_{\bf k} \right] \gamma_0$, Eq.\
(\ref{quasiparticleprop}) can be brought in the form
\begin{equation}
G^\pm(K) = \sum_{e= \pm} \frac{k_0 \mp (\mu - ek)}{
k_0^2 - [ \epsilon_{\bf k}^{e} ]^2 } \, 
\Lambda^{\pm e}_{\bf k} \, \gamma_0 \,\,,
\end{equation}
which should be compared with Eq.\ (\ref{freequarkprop}).
Obviously, all that has changed is that the free quark excitation
energies (\ref{freequarkenergy}) have been replaced by the
quasiparticle excitation energies (\ref{quasiparticleenergy}),
$\epsilon_{{\bf k} 0}^e \rightarrow \epsilon_{\bf k}^e
\equiv \epsilon_{\bf k}^e (\phi^e)$.

After realizing this, by comparison with Eqs.\ (\ref{mixedprops})
one can immediately write down the mixed representation for the
quasiparticle propagators,
\begin{mathletters} \label{mixedquasiparticleprops}
\begin{eqnarray}
G^+(\tau,{\bf k}) & = & - \sum_{e = \pm} 
\Lambda^{e}_{\bf k}\, \gamma_0 \left\{ \frac{}{}\! (1-n_{\bf k}^e)\,
\left[ \theta(\tau) - N(\epsilon^e_{\bf k}) \right]
\exp(-\epsilon_{\bf k}^e \tau) 
-n_{\bf k}^e\, \left[ \theta(-\tau) - N(\epsilon^e_{\bf k}) \right] 
\exp(\epsilon_{\bf k}^e \tau) \right\} ,\\
G^-(\tau,{\bf k}) & = & - \sum_{e = \pm} 
\gamma_0\, \Lambda^e_{\bf k} \left\{ \frac{}{}\! n_{\bf k}^e\,
\left[ \theta(\tau) - N(\epsilon^e_{\bf k}) \right] 
\exp(-\epsilon_{\bf k}^e \tau) 
-(1-n_{\bf k}^e)\, \left[ \theta(-\tau) - N(\epsilon^e_{\bf k}) 
\right]  \exp(\epsilon_{\bf k}^e \tau) \right\}\,\, .
\end{eqnarray}
\end{mathletters}
Here, 
\begin{equation}
n_{\bf k}^e \equiv \frac{\epsilon_{\bf k}^e + \mu - ek}{2 \,
\epsilon_{\bf k}^e} 
\end{equation}
are the occupation numbers for quasiparticles ($e = +1$) or quasi-antiparticles
($e = -1$) at zero temperature \cite{prscalar}. Consequently,
$1-n_{\bf k}^e$ are the occupation numbers for quasiparticle holes
or quasi-antiparticle holes. Due to the presence of a gap $\phi^e$ in
the quasiparticle excitation spectrum, these occupation numbers are no longer
simple theta functions in momentum space, as in the noninteracting case;
the theta functions become ``smeared'' over a range $\sim \phi^e$ around
the Fermi surface (cf.\ Fig.\ 2 in \cite{prscalar}).
The relations (\ref{chargeconjprop}) and (\ref{KMS}) are also
fulfilled by $G^\pm (\tau,{\bf k})$.

From a comparison of Eqs.\ (\ref{mixedquasiparticleprops}) and 
(\ref{mixedprops}), one can immediately deduce from Eq.\
(\ref{Pi}) the result for the traces
${\rm Tr}_s \left[\gamma^\mu\, G^\pm(K)\, \gamma^\nu G^\pm(K-P) \right]$,
${\rm Tr}_s \left[\gamma^\mu\, G_0^\pm(K)\, \gamma^\nu G^\pm(K-P) \right]$,
or
${\rm Tr}_s \left[\gamma^\mu\, G^\pm(K)\, \gamma^\nu G_0^\pm(K-P) \right]$.
All one has to do is replace
\begin{equation}
\epsilon^0_i \rightarrow \epsilon_i \equiv \epsilon^{e_i}_{{\bf k}_i} \;\;\; ,
\;\;\;\;
n^0_i \rightarrow n_i \equiv n^{e_i}_{{\bf k}_i} \;\;\; , \;\;\;\;
N^0_i \rightarrow N_i \equiv N(\epsilon_i) \,\, ,
\end{equation}
when a propagator $G^\pm$ occurs in place of $G_0^\pm$.

One also needs a mixed representation for the off-diagonal components
of ${\cal S}(K)$. First, write $\Xi^\pm(K)$, Eq.\ (\ref{offdiagonal}),
in terms of projectors,
\begin{equation}
\Xi^+(K) = - \sum_{h = r,\ell} \sum_{e=\pm}
\frac{\phi^e_h(K)}{k_0^2 - [ \epsilon_{\bf k}^e]^2} \, {\cal P}_{-h} \,
\Lambda^{-e}_{\bf k} \;\;\; , \;\;\;\;
\Xi^-(K) = - \sum_{h = r,\ell} \sum_{e=\pm}
\frac{\left[\phi^e_h(K)\right]^*}{k_0^2 - [ \epsilon_{\bf k}^e]^2} \, 
{\cal P}_h \, \Lambda^e_{\bf k} \,\, .
\end{equation}
As in \cite{prqcd}, assume that $\phi^e_h(k_0)$ has no
poles or cuts in the complex $k_0$-plane and that
$\phi^e_h(k_0) = \phi^e_h(-k_0)$. In this case, one
obtains the mixed representations
\begin{mathletters} \label{mixedoffdiagonal}
\begin{eqnarray}
\Xi^+(\tau,{\bf k}) & = & \sum_{h=r,\ell} \sum_{e=\pm}
{\cal P}_{-h}\, \Lambda^{-e}_{\bf k} \;
\frac{\phi_h^e(\epsilon_{\bf k}^e,{\bf k})}{2\, \epsilon_{\bf k}^e}\,
\left\{ \frac{}{}\! \left[\theta(\tau) - N(\epsilon_{\bf k}^e) \right]
\, \exp ( - \epsilon_{\bf k}^e \tau) + 
\left[\theta(-\tau) - N(\epsilon_{\bf k}^e) \right]
\, \exp ( \epsilon_{\bf k}^e \tau) \right\} \,\, , \\
\Xi^-(\tau,{\bf k}) & = & \sum_{h=r,\ell} \sum_{e=\pm}
{\cal P}_h\, \Lambda^e_{\bf k} \;
\frac{\left[\phi_h^e(\epsilon_{\bf k}^e,{\bf k})\right]^*}{2\, 
\epsilon_{\bf k}^e}\,
\left\{ \frac{}{}\! \left[\theta(\tau) - N(\epsilon_{\bf k}^e) \right]
\, \exp ( - \epsilon_{\bf k}^e \tau) + 
\left[\theta(-\tau) - N(\epsilon_{\bf k}^e) \right]
\, \exp ( \epsilon_{\bf k}^e \tau) \right\} \,\, .
\end{eqnarray}
\end{mathletters}
Note that the energy in the gap functions $\phi_h^e$ is on the 
quasiparticle mass shell, $k_0 \equiv \pm \epsilon_{\bf k}^e$.

The traces in Eqs.\ (\ref{Pisupercond})
involving $\Xi^\pm$ are now straightforwardly computed as
\begin{eqnarray}
\lefteqn{
T \sum_{k_0} {\rm Tr}_s \left[ \gamma^\mu\, \Xi^\mp(K_1) \, \gamma^\nu \,
\Xi^\pm(K_2) \right] = \sum_{e_1,e_2 =\pm}
{\cal U}_\pm^{\mu \nu}({\bf k}_1,{\bf k}_2)\, 
\frac{\phi_1 \, \phi_2}{4 \, \epsilon_1\, \epsilon_2}} \nonumber \\
&  \times  &
\left[ \left( \frac{1}{p_0 + \epsilon_1 + \epsilon_2} - \frac{1}{
p_0 - \epsilon_1 - \epsilon_2} \right) \left( 1- N_1 - N_2 \right)
- \left( \frac{1}{p_0 - \epsilon_1 + \epsilon_2} - \frac{1}{p_0 + \epsilon_1
- \epsilon_2} \right) \left( N_1 - N_2 \right) \right] \,\, ,
\end{eqnarray}
where $K_1 \equiv K$, $K_2 \equiv K-P$, as before, while
\begin{equation}
\phi_i \equiv \phi^{e_i}(\epsilon_i,{\bf k}_i)\,\, ,
\end{equation}
and
\begin{equation}
{\cal U}_\pm^{\mu \nu} ({\bf k}_1,{\bf k}_2) \equiv
{\rm Tr}_s \left[ \gamma^\mu \, \Lambda^{\pm e_1}_{{\bf k}_1} \,
\gamma^\nu \, \Lambda^{\mp e_2}_{{\bf k}_2} \right] \,\, .
\end{equation}
On account of ${\cal P}_h \, \gamma^\mu = \gamma^\mu \,
{\cal P}_{-h}$ and ${\cal P}_r\, {\cal P}_\ell = 0$, the sum over
chiralities $h_1$ and $h_2$ originating
from the mixed representations (\ref{mixedoffdiagonal})
could be performed trivially.
\newpage
Putting everything together, the self-energy for gluons of color 1, 2, and 3
is
\begin{mathletters}
\begin{eqnarray}
\lefteqn{ \Pi^{\mu \nu}_{11} (P) =  - \frac{1}{4}\, g^2\, N_f
\, \int \frac{d^3{\bf k}}{(2 \pi)^3} \sum_{e_1,e_2 = \pm} 
 \left\{ \frac{}{}\!
{\cal T}_+^{\mu \nu}({\bf k}_1,{\bf k}_2) \right. } \nonumber \\
& \times &  \left[ \left( \frac{n_1 \, (1-n_2)}{p_0 + \epsilon_1
+ \epsilon_2} - \frac{(1-n_1)\, n_2}{p_0 - \epsilon_1
- \epsilon_2} \right) (1-N_1 - N_2) 
+ \left( \frac{(1-n_1)\, (1-n_2)}{p_0 - \epsilon_1
+ \epsilon_2} - \frac{n_1\, n_2}{p_0 + \epsilon_1
- \epsilon_2} \right) (N_1 - N_2) \right]  \nonumber \\
&   & \hspace*{5cm} + \;\;
 {\cal T}_-^{\mu \nu}({\bf k}_1,{\bf k}_2) \nonumber \\
& \times & \left[ \left( \frac{(1-n_1)\, n_2}{p_0 + \epsilon_1
+ \epsilon_2} - \frac{n_1\, (1- n_2)}{p_0 - \epsilon_1
- \epsilon_2} \right) (1-N_1 - N_2) 
+ \left( \frac{n_1 \, n_2}{p_0 - \epsilon_1
+ \epsilon_2} - \frac{(1-n_1)\,(1- n_2)}{p_0 + \epsilon_1
- \epsilon_2} \right) (N_1 - N_2) \right]  \nonumber \\
&   & \hspace*{5cm} - \;\left[ {\cal U}^{\mu \nu}_+({\bf k}_1, {\bf k}_2) + 
{\cal U}^{\mu \nu}_- ({\bf k}_1, {\bf k}_2) \right] \, 
\frac{\phi_1\, \phi_2}{4 \, \epsilon_1\, \epsilon_2} \nonumber \\
& \times &  \left.
\left[ \left( \frac{1}{p_0 + \epsilon_1 + \epsilon_2} - \frac{1}{
p_0 - \epsilon_1 - \epsilon_2} \right) \left( 1- N_1 - N_2 \right)
- \left( \frac{1}{p_0 - \epsilon_1 + \epsilon_2} - \frac{1}{p_0 + \epsilon_1
- \epsilon_2} \right) \left( N_1 - N_2 \right) \right] \right\}
\,\, , \label{Pi11b}
\end{eqnarray}
for gluon colors 4 and 6,
\begin{eqnarray}
\lefteqn{ \Pi^{\mu \nu}_{44} (P)  =  - \frac{1}{4}\, g^2\, N_f
\, \int \frac{d^3{\bf k}}{(2 \pi)^3} \sum_{e_1,e_2 = \pm}  
 \left\{ \frac{}{}\! 
{\cal T}_+^{\mu \nu}({\bf k}_1,{\bf k}_2)\right.} \nonumber \\
& \times & \left[ \left( \frac{n^0_1 \, (1-n_2)}{p_0 + \epsilon^0_1
+ \epsilon_2} - \frac{(1-n^0_1)\, n_2}{p_0 - \epsilon^0_1
- \epsilon_2} \right) (1-N^0_1 - N_2) +
\left( \frac{(1-n^0_1)\, (1-n_2)}{p_0 - \epsilon^0_1
+ \epsilon_2} - \frac{n^0_1\, n_2}{p_0 + \epsilon^0_1
- \epsilon_2} \right) (N^0_1 - N_2)  \right] \nonumber \\
&   & \hspace*{5.1cm} +  \; {\cal T}_-^{\mu \nu}({\bf k}_1,{\bf k}_2)
\nonumber \\
& \times & \left.\!\!  \left[ 
\left( \frac{(1-n_1)\, n^0_2}{p_0 + \epsilon_1
+ \epsilon^0_2} - \frac{n_1\, (1- n^0_2)}{p_0 - \epsilon_1
- \epsilon^0_2} \right)\! (1-N_1 - N^0_2) +
\left( \frac{n_1 \, n^0_2}{p_0 - \epsilon_1
+ \epsilon^0_2} - \frac{(1-n_1)\,(1- n^0_2)}{p_0 + \epsilon_1
- \epsilon^0_2} \right)\! (N_1 - N^0_2) 
\right] \right\} ,\!\! \label{Pi44b}
\end{eqnarray}
for gluon colors 5 and 7,
\begin{eqnarray}
\lefteqn{ \Pi^{\mu \nu}_{55} (P)  =  - \frac{1}{4}\, g^2\, N_f
\, \int \frac{d^3{\bf k}}{(2 \pi)^3} \sum_{e_1,e_2 = \pm}  
 \left\{ \frac{}{}\! 
{\cal T}_+^{\mu \nu}({\bf k}_1,{\bf k}_2)\right.} \nonumber \\
& \times & \left[ \left( \frac{n_1 \, (1-n^0_2)}{p_0 + \epsilon_1
+ \epsilon^0_2} - \frac{(1-n_1)\, n^0_2}{p_0 - \epsilon_1
- \epsilon^0_2} \right) (1-N_1 - N^0_2) +
 \left( \frac{(1-n_1)\, (1-n^0_2)}{p_0 - \epsilon_1
+ \epsilon^0_2} - \frac{n_1\, n^0_2}{p_0 + \epsilon_1
- \epsilon^0_2} \right) (N_1 - N^0_2) \right] \nonumber \\
&   & \hspace*{5.1cm} +  \; {\cal T}_-^{\mu \nu}({\bf k}_1,{\bf k}_2)
\nonumber \\
& \times & \left. \!\! \left[ \left( \frac{(1-n^0_1)\, n_2}{p_0 + \epsilon^0_1
+ \epsilon_2} - \frac{n^0_1\, (1- n_2)}{p_0 - \epsilon^0_1
- \epsilon_2} \right) \! (1-N^0_1 - N_2) +
\left( \frac{n^0_1 \, n_2}{p_0 - \epsilon^0_1
+ \epsilon_2} - \frac{(1-n^0_1)\,(1- n_2)}{p_0 + \epsilon^0_1
- \epsilon_2} \right) \! (N^0_1 - N_2) 
\right] \right\} ,\!\! \label{Pi55b}
\end{eqnarray}
and for gluon color 8
\begin{eqnarray}
\lefteqn{\tilde{\Pi}^{\mu \nu} (P)  =  - \frac{1}{4}\, g^2\, N_f
\, \int \frac{d^3{\bf k}}{(2 \pi)^3} \sum_{e_1,e_2 = \pm}  
 \left\{ \frac{}{}\! {\cal T}_+^{\mu \nu}({\bf k}_1,{\bf k}_2) \right.}
\nonumber \\
& \times &  \left[ \left( \frac{n_1 \, (1-n_2)}{p_0 + \epsilon_1
+ \epsilon_2} - \frac{(1-n_1)\, n_2}{p_0 - \epsilon_1
- \epsilon_2} \right) (1-N_1 - N_2) 
+ \left( \frac{(1-n_1)\, (1-n_2)}{p_0 - \epsilon_1
+ \epsilon_2} - \frac{n_1\, n_2}{p_0 + \epsilon_1
- \epsilon_2} \right) (N_1 - N_2) \right]  \nonumber \\
&   & \hspace*{5cm} + \;\;
 {\cal T}_-^{\mu \nu}({\bf k}_1,{\bf k}_2) \nonumber \\
& \times & \left[ \left( \frac{(1-n_1)\, n_2}{p_0 + \epsilon_1
+ \epsilon_2} - \frac{n_1\, (1- n_2)}{p_0 - \epsilon_1
- \epsilon_2} \right) (1-N_1 - N_2) 
+ \left( \frac{n_1 \, n_2}{p_0 - \epsilon_1
+ \epsilon_2} - \frac{(1-n_1)\,(1- n_2)}{p_0 + \epsilon_1
- \epsilon_2} \right) (N_1 - N_2) \right]  \nonumber \\
&   & \hspace*{5cm} + \;\left[ {\cal U}^{\mu \nu}_+({\bf k}_1, {\bf k}_2) + 
{\cal U}^{\mu \nu}_- ({\bf k}_1, {\bf k}_2) \right] \, 
\frac{\phi_1\, \phi_2}{4 \, \epsilon_1\, \epsilon_2} \nonumber \\
& \times &  \left.
\left[ \left( \frac{1}{p_0 + \epsilon_1 + \epsilon_2} - \frac{1}{
p_0 - \epsilon_1 - \epsilon_2} \right) \left( 1- N_1 - N_2 \right)
- \left( \frac{1}{p_0 - \epsilon_1 + \epsilon_2} - \frac{1}{p_0 + \epsilon_1
- \epsilon_2} \right) \left( N_1 - N_2 \right) \right] \right\}
\,\, . \label{Pi88b}
\end{eqnarray}
\end{mathletters}

\subsection{Trace over spinor space}
 
The traces ${\cal T}_\pm^{\mu \nu}$ have been computed in
Sec.\ \ref{iiie}. What remains to be done is to compute
${\cal U}_\pm^{\mu \nu}$. One finds
\begin{mathletters} \label{U}
\begin{eqnarray}
{\cal U}_\pm^{00} & = & {\cal T}_\pm^{00} \,\, , \label{Ua} \\
{\cal U}_\pm^{0i} = - {\cal U}_\pm^{i0} & = & - {\cal T}_\pm^{0i} 
\;\;\; , \;\;\;\; i = x,y,x \,\, , \label{Ub} \\
{\cal U}_\pm^{ij} & = & - {\cal T}_\pm^{ij} \;\;\; , \;\;\;\;
i,j = x,y,z \,\,. \label{Uc}
\end{eqnarray}
\end{mathletters}
In the following, the results for the different components of
the gluon self-energy in the superconducting phase will be collected.

\subsection{Gluons of color 1, 2, and 3}

\noindent
\underline{(i) $\mu = \nu =0$:}\hspace*{0.5cm} Defining
\begin{equation}
\xi_i \equiv e_i\, k_i - \mu\,\,,
\end{equation}
the self-energy of electric gluons of color 1, 2, and 3 is determined
from Eqs.\ (\ref{traces2}), (\ref{Pi11b}), and (\ref{Ua}) as
\begin{mathletters} \label{Pi11all}
\begin{eqnarray}
\Pi^{00}_{11}(P) & = & - \frac{1}{4}\, g^2 \, N_f \int 
\frac{d^3{\bf k}}{(2 \pi)^3}  \sum_{e_1,e_2 = \pm} (1 + e_1 e_2\,
\hat{\bf k}_1 \cdot \hat{\bf k}_2) \nonumber \\
&  & \times \left[ \left( \frac{1}{p_0 + \epsilon_1 + \epsilon_2}
- \frac{1}{p_0 - \epsilon_1 - \epsilon_2} \right) 
\left( 1- N_1 - N_2 \right) \, 
\frac{ \epsilon_1 \, \epsilon_2 - \xi_1 \, \xi_2 - \phi_1\, \phi_2}{
2 \, \epsilon_1 \, \epsilon_2} \right. \nonumber \\
&  & \left.\;+ \, \left( \frac{1}{p_0 - \epsilon_1 + \epsilon_2}
- \frac{1}{p_0 + \epsilon_1 - \epsilon_2} \right) 
\left(N_1 - N_2 \right) \, 
\frac{ \epsilon_1 \, \epsilon_2 + \xi_1\, \xi_2 + \phi_1 \, \phi_2}{
2 \, \epsilon_1\,  \epsilon_2} \right] \,\, . \label{Pi1100}
\end{eqnarray}
\\ ~~ \\
\noindent
\underline{(ii) $\mu = 0, \, \nu = i$:} \hspace*{0.5cm}
For the $(0i)$-components of the 
self-energy of gluons with colors 1, 2, or 3 one obtains
\begin{eqnarray}
\Pi^{0i}_{11}(P) & = & - \frac{1}{4}\, g^2 \, N_f \int 
\frac{d^3{\bf k}}{(2 \pi)^3}  \sum_{e_1,e_2 = \pm} ( e_1 \,
\hat{k}_1^i + e_2\, \hat{k}_2^i) \nonumber \\
&  & \times \left[ \left( \frac{1}{p_0 + \epsilon_1 + \epsilon_2}
+ \frac{1}{p_0 - \epsilon_1 - \epsilon_2} \right) 
\left( 1- N_1 - N_2 \right) \, \left( \frac{\xi_2}{2\, \epsilon_2}
- \frac{\xi_1}{2\, \epsilon_1} \right) \right. \nonumber \\
&  & \left.\;+ \, \left( \frac{1}{p_0 - \epsilon_1 + \epsilon_2}
+ \frac{1}{p_0 + \epsilon_1 - \epsilon_2} \right) 
\left(N_1 - N_2 \right) \, \left( \frac{\xi_1}{2\, \epsilon_1} +
\frac{ \xi_2}{2\, \epsilon_2} \right)\right] \,\, . \label{Pi110i}
\end{eqnarray}
\\ ~~ \\
\noindent
\underline{(iii) $\mu = i,\, \nu =j$:}\hspace*{0.5cm}
The self-energy of magnetic gluons of colors 1, 2, and 3 is
\begin{eqnarray}
\Pi^{ij}_{11}(P) & = & - \frac{1}{4}\, g^2 \, N_f \int 
\frac{d^3{\bf k}}{(2 \pi)^3}  \sum_{e_1,e_2 = \pm} 
\left[ \delta^{ij} \left(1 - e_1 e_2\,\hat{\bf k}_1 \cdot \hat{\bf k}_2\right) 
+ e_1 e_2 \left( \hat{k}_1^i \, \hat{k}_2^j + \hat{k}_1^j \,
\hat{k}_2^i \right) \right] \nonumber \\
&  & \times \left[ \left( \frac{1}{p_0 + \epsilon_1 + \epsilon_2}
- \frac{1}{p_0 - \epsilon_1 - \epsilon_2} \right) 
\left( 1- N_1 - N_2 \right) \, 
\frac{ \epsilon_1 \, \epsilon_2 - \xi_1 \, \xi_2 + \phi_1\, \phi_2}{
2 \, \epsilon_1 \, \epsilon_2} \right. \nonumber \\
&  & \left.\;+ \, \left( \frac{1}{p_0 - \epsilon_1 + \epsilon_2}
- \frac{1}{p_0 + \epsilon_1 - \epsilon_2} \right) 
\left(N_1 - N_2 \right) \, 
\frac{ \epsilon_1 \, \epsilon_2 + \xi_1\, \xi_2 - \phi_1 \, \phi_2}{
2 \, \epsilon_1\,  \epsilon_2} \right] \,\, . \label{Pi11ij}
\end{eqnarray}
\end{mathletters}

\subsection{Gluons of color 4 and 6}

\noindent
\underline{(i) $\mu = \nu = 0$:}\hspace*{0.5cm}
Using the symmetry of Eq.\ (\ref{Pi44b}) under 
${\bf k}_1 \leftrightarrow - {\bf k}_2$, $e_1 \leftrightarrow e_2$,
the self-energy of electric gluons of colors 4 and 6 can be written as
\begin{mathletters} \label{Pi44all}
\begin{eqnarray}
\lefteqn{\Pi^{00}_{44} (P)  =  - \frac{1}{2}\, g^2\, N_f
\, \int \frac{d^3{\bf k}}{(2 \pi)^3} \sum_{e_1,e_2 = \pm}  
(1 + e_1 e_2\, \hat{\bf k}_1 \cdot \hat{\bf k}_2) }\nonumber \\ 
& \times & \! \left[ \left( \frac{n^0_1 \, (1-n_2)}{p_0 + \epsilon^0_1
+ \epsilon_2} - \frac{(1-n^0_1)\, n_2}{p_0 - \epsilon^0_1
- \epsilon_2} \right) \! (1-N^0_1 - N_2) +
\left( \frac{(1-n^0_1)\, (1-n_2)}{p_0 - \epsilon^0_1
+ \epsilon_2} - \frac{n^0_1\, n_2}{p_0 + \epsilon^0_1
- \epsilon_2} \right)\! (N^0_1 - N_2)  \right] . \! \label{Pi4400}
\end{eqnarray}
\\ ~~ \\
\noindent
\underline{(ii) $\mu = 0, \, \nu = i$:} \hspace*{0.5cm}
The same symmetry arguments lead to
\begin{eqnarray}
\lefteqn{\Pi^{0i}_{44} (P)  =  - \frac{1}{2}\, g^2\, N_f
\, \int \frac{d^3{\bf k}}{(2 \pi)^3} \sum_{e_1,e_2 = \pm}  
(e_1\, \hat{k}_1^i +  e_2\, \hat{k}_2^i) }\nonumber \\ 
& \times & \! \left[ \left( \frac{n^0_1 \, (1-n_2)}{p_0 + \epsilon^0_1
+ \epsilon_2} - \frac{(1-n^0_1)\, n_2}{p_0 - \epsilon^0_1
- \epsilon_2} \right) \! (1-N^0_1 - N_2) +
\left( \frac{(1-n^0_1)\, (1-n_2)}{p_0 - \epsilon^0_1
+ \epsilon_2} - \frac{n^0_1\, n_2}{p_0 + \epsilon^0_1
- \epsilon_2} \right)\! (N^0_1 - N_2)  \right] . \! \label{Pi440i}
\end{eqnarray}
\\ ~~ \\
\noindent
\underline{(iii) $\mu = i,\, \nu = j$:}\hspace*{0.5cm}
For the self-energy of magnetic gluons of color 4 and 6 one obtains
\begin{eqnarray}
\lefteqn{\Pi^{ij}_{44} (P)  =  - \frac{1}{2}\, g^2\, N_f
\, \int \frac{d^3{\bf k}}{(2 \pi)^3} \sum_{e_1,e_2 = \pm}  
\left[ \delta^{ij} (1-e_1 e_2\, \hat{\bf k}_1 \cdot \hat{\bf k}_2) 
 + e_1 e_2 \, \left( \hat{k}_1^i\, \hat{k}_2^j + \hat{k}_1^j\, \hat{k}_2^i
\right) \right] }\nonumber \\ 
& \times & \left[ \left( \frac{n^0_1 \, (1-n_2)}{p_0 + \epsilon^0_1
+ \epsilon_2} - \frac{(1-n^0_1)\, n_2}{p_0 - \epsilon^0_1
- \epsilon_2} \right) \! (1-N^0_1 - N_2) +
\left( \frac{(1-n^0_1)\, (1-n_2)}{p_0 - \epsilon^0_1
+ \epsilon_2} - \frac{n^0_1\, n_2}{p_0 + \epsilon^0_1
- \epsilon_2} \right)\! (N^0_1 - N_2)  \right] . \! \label{Pi44ij}
\end{eqnarray}
\end{mathletters}

\subsection{Gluons of color 5 and 7}

\noindent
\underline{(i) $\mu = \nu = 0$:} \hspace*{0.5cm}
Again using the symmetry of Eq.\ (\ref{Pi55b}) under 
${\bf k}_1 \leftrightarrow - {\bf k}_2$, $e_1 \leftrightarrow e_2$,
the self-energy of electric gluons of colors 5 and 7 can be written as
\begin{mathletters} \label{Pi55all}
\begin{eqnarray}
\lefteqn{\Pi^{00}_{55} (P)  =  - \frac{1}{2}\, g^2\, N_f
\, \int \frac{d^3{\bf k}}{(2 \pi)^3} \sum_{e_1,e_2 = \pm}  
(1 + e_1 e_2\, \hat{\bf k}_1 \cdot \hat{\bf k}_2) }\nonumber \\ 
& \times & \! \left[ \left( \frac{(1-n^0_1) \, n_2}{p_0 + \epsilon^0_1
+ \epsilon_2} - \frac{n^0_1\,(1- n_2)}{p_0 - \epsilon^0_1
- \epsilon_2} \right) \! (1-N^0_1 - N_2) +
\left( \frac{n^0_1\, n_2}{p_0 - \epsilon^0_1
+ \epsilon_2} - \frac{(1-n^0_1)\,(1- n_2)}{p_0 + \epsilon^0_1
- \epsilon_2} \right)\! (N^0_1 - N_2)  \right] . \! \label{Pi5500}
\end{eqnarray}
\\ ~~ \\
\noindent
\underline{(ii) $\mu = 0,\, \nu= i$:}\hspace*{0.5cm}
The $(0i)$-components are
\begin{eqnarray}
\lefteqn{\Pi^{0i}_{55} (P)  =   \frac{1}{2}\, g^2\, N_f
\, \int \frac{d^3{\bf k}}{(2 \pi)^3} \sum_{e_1,e_2 = \pm}  
(e_1\, \hat{k}_1^i +  e_2\, \hat{k}_2^i) }\nonumber \\ 
& \times & \! \left[ \left( \frac{(1-n^0_1) \, n_2}{p_0 + \epsilon^0_1
+ \epsilon_2} - \frac{n^0_1\,(1- n_2)}{p_0 - \epsilon^0_1
- \epsilon_2} \right) \! (1-N^0_1 - N_2) +
\left( \frac{n^0_1\, n_2}{p_0 - \epsilon^0_1
+ \epsilon_2} - \frac{(1-n^0_1)\, (1-n_2)}{p_0 + \epsilon^0_1
- \epsilon_2} \right)\! (N^0_1 - N_2)  \right] . \! \label{Pi550i}
\end{eqnarray}
\\ ~~ \\
\noindent
\underline{(iii) $\mu = i,\, \nu = j$:}\hspace*{0.5cm}
For the self-energy of magnetic gluons of color 5 and 7 one gets
\begin{eqnarray}
\lefteqn{\Pi^{ij}_{55} (P)  =  - \frac{1}{2}\, g^2\, N_f
\, \int \frac{d^3{\bf k}}{(2 \pi)^3} \sum_{e_1,e_2 = \pm}  
\left[ \delta^{ij} (1-e_1 e_2\, \hat{\bf k}_1 \cdot \hat{\bf k}_2) 
 + e_1 e_2 \, \left( \hat{k}_1^i\, \hat{k}_2^j + \hat{k}_1^j\, \hat{k}_2^i
\right) \right] }\nonumber \\ 
& \times & \left[ \left( \frac{(1-n^0_1) \, n_2}{p_0 + \epsilon^0_1
+ \epsilon_2} - \frac{n^0_1\,(1- n_2)}{p_0 - \epsilon^0_1
- \epsilon_2} \right) \! (1-N^0_1 - N_2) +
\left( \frac{n^0_1\, n_2}{p_0 - \epsilon^0_1
+ \epsilon_2} - \frac{(1-n^0_1)\, (1-n_2)}{p_0 + \epsilon^0_1
- \epsilon_2} \right)\! (N^0_1 - N_2)  \right] . \! \label{Pi55ij}
\end{eqnarray}
\end{mathletters}

\subsection{Gluons of color 8}

\noindent
\underline{(i) $\mu = \nu = 0$:}\hspace*{0.5cm}
For $\tilde{\Pi}^{00}$ one obtains
\begin{mathletters} \label{Pi88all}
\begin{eqnarray}
\tilde{\Pi}^{00}(P) & = & - \frac{1}{4}\, g^2 \, N_f \int 
\frac{d^3{\bf k}}{(2 \pi)^3}  \sum_{e_1,e_2 = \pm} (1 + e_1 e_2\,
\hat{\bf k}_1 \cdot \hat{\bf k}_2) \nonumber \\
&  & \times \left[ \left( \frac{1}{p_0 + \epsilon_1 + \epsilon_2}
- \frac{1}{p_0 - \epsilon_1 - \epsilon_2} \right) 
\left( 1- N_1 - N_2 \right) \, 
\frac{ \epsilon_1 \, \epsilon_2 - \xi_1 \, \xi_2 + \phi_1\, \phi_2}{
2 \, \epsilon_1 \, \epsilon_2} \right. \nonumber \\
&  & \left.\;+ \, \left( \frac{1}{p_0 - \epsilon_1 + \epsilon_2}
- \frac{1}{p_0 + \epsilon_1 - \epsilon_2} \right) 
\left(N_1 - N_2 \right) \, 
\frac{ \epsilon_1 \, \epsilon_2 + \xi_1\, \xi_2 - \phi_1 \, \phi_2}{
2 \, \epsilon_1\,  \epsilon_2} \right] \,\, . \label{Pi8800}
\end{eqnarray}
\\ ~~ \\
\noindent
\underline{(ii) $\mu = 0,\, \nu = i$:}\hspace*{0.5cm}
For $\tilde{\Pi}^{0i}$ one simply has
\begin{equation} \label{Pi880i}
\tilde{\Pi}^{0i}(P) \equiv \Pi^{0i}_{11}(P)\,\,.
\end{equation}
\\ ~~ \\
\noindent
\underline{(iii) $\mu = i, \, \nu = j$:}\hspace*{0.5cm}
The magnetic components $\tilde{\Pi}^{ij}$ are
\begin{eqnarray}
\tilde{\Pi}^{ij}(P) & = & - \frac{1}{4}\, g^2 \, N_f \int 
\frac{d^3{\bf k}}{(2 \pi)^3}  \sum_{e_1,e_2 = \pm} 
\left[ \delta^{ij} \left(1 - e_1 e_2\,\hat{\bf k}_1 \cdot \hat{\bf k}_2\right) 
+ e_1 e_2 \left( \hat{k}_1^i \, \hat{k}_2^j + \hat{k}_1^j \,
\hat{k}_2^i \right) \right] \nonumber \\
&  & \times \left[ \left( \frac{1}{p_0 + \epsilon_1 + \epsilon_2}
- \frac{1}{p_0 - \epsilon_1 - \epsilon_2} \right) 
\left( 1- N_1 - N_2 \right) \, 
\frac{ \epsilon_1 \, \epsilon_2 - \xi_1 \, \xi_2 - \phi_1\, \phi_2}{
2 \, \epsilon_1 \, \epsilon_2} \right. \nonumber \\
&  & \left.\;+ \, \left( \frac{1}{p_0 - \epsilon_1 + \epsilon_2}
- \frac{1}{p_0 + \epsilon_1 - \epsilon_2} \right) 
\left(N_1 - N_2 \right) \, 
\frac{ \epsilon_1 \, \epsilon_2 + \xi_1\, \xi_2 + \phi_1 \, \phi_2}{
2 \, \epsilon_1\,  \epsilon_2} \right] \,\, . \label{Pi88ij}
\end{eqnarray}
\end{mathletters}
Equations (\ref{Pi11all}) -- (\ref{Pi88all}) are the central
result of this work. Starting from these equations, one can derive explicit
expressions for the gluon self-energy in a two-flavor color superconductor
for arbitrary $p_0$ and ${\bf p}$. As a first step, in the remainder of 
this work I compute the color-electric (Debye) screening mass, as well as
the color-magnetic (Meissner) mass. 
These are obtained from the gluon self-energy
in the static limit, $p_0 = 0$, for $p \rightarrow 0$.
Then I compute the self-energy for $p_0 = 0$, but $p \gg \phi_0$.

\section{Debye screening and Meissner effect} \label{v}

In the following, I shall always assume that antiparticle gaps
are small, $\phi^- \simeq 0$, and consequently that
\begin{equation} \label{antiparticleapprox}
\epsilon^-_{\bf k} \simeq \epsilon^-_{{\bf k}0} \;\;\; , \;\;\;\;
n^-_{\bf k} \simeq n^-_{{\bf k}0} \simeq 1
\;\;\; , \;\;\;\;
N(\epsilon^-_{\bf k}) \simeq 0 \,\, .
\end{equation}
Therefore, thermal antiparticle occupation numbers and their derivatives
will be neglected. As in the previous section, the different
color sectors will be discussed separately.

\subsection{Gluons with colors 1, 2, and 3}

\noindent
\underline{(i) $\mu = \nu =0$:}\hspace*{0.5cm}
I show several calculational steps in greater detail to illustrate
the main approximations used throughout the following.
For $p_0 = 0$, $p \rightarrow 0$, ${\bf k}_2 \rightarrow {\bf k}_1
\equiv {\bf k}$, and only particle-particle ($e_1 = e_2 = +1$), or
antiparticle-antiparticle ($e_1 = e_2 = -1$) excitations contribute
in the sum over $e_1$ and $e_2$ in (\ref{Pi1100}). This is very
similar to what happens in the HDL limit, cf.\ Sec.\ \ref{HDL}.
Furthermore
\begin{equation}
\frac{ \epsilon_1 \, \epsilon_2 - \xi_1 \, \xi_2 - \phi_1\, \phi_2}{
2 \, \epsilon_1 \, \epsilon_2} \rightarrow 0 \;\;\; ,\;\;\;\;
\frac{ \epsilon_1 \, \epsilon_2 + \xi_1 \, \xi_2 + \phi_1\, \phi_2}{
2 \, \epsilon_1 \, \epsilon_2} \rightarrow 1 \,\, .
\end{equation}
In the limit ${\bf k}_2 \rightarrow {\bf k}_1$,
$(N_1 - N_2)/(\epsilon_1 - \epsilon_2) \rightarrow
dN/d\epsilon$, and neglecting the variation of $N(\epsilon^-_{\bf k})$,
as discussed above, one obtains
\begin{equation}
\Pi^{00}_{11}(0) \simeq  \frac{g^2 \, N_f}{2 \pi^2} \int_0^\infty 
dk \, k^2 \, \frac{d N(\epsilon^+_{\bf k})}{d \epsilon^+_{\bf k}} \,\, .
\end{equation}
As the thermal occupation number varies appreciably only close to
the Fermi surface, it is permissible to approximate $k^2 \simeq \mu^2$,
and to restrict the $k$ integration to the region $0 \leq k \leq 2 \mu$.
Introducing the variable 
\begin{equation} \label{xi}
\xi \equiv k - \mu \,\, ,
\end{equation}
one obtains with Eq.\ (\ref{gluonmass})
\begin{equation}
\Pi^{00}_{11}(0) \simeq  - 3\, m_g^2 \int_0^{\mu} 
\frac{d\xi}{2T} \, \frac{1}{\cosh^2 \left(\sqrt{\xi^2 + \phi^2}/2T \right)} 
\,\, .
\end{equation}
Now change variables to $\zeta \equiv \xi/2T$, and remembering that
$\mu \gg \phi \sim T$, send the upper limit of the integral to
infinity,
\begin{equation}
\Pi^{00}_{11}(0) \simeq  - 3\, m_g^2 \int_0^\infty 
d\zeta \, \frac{1}{\cosh^2 \sqrt{\zeta^2 + (\phi/2T)^2}} \,\, .
\end{equation}
This expression has two interesting limits. For $T \rightarrow 0$,
the integrand becomes zero everywhere, and
\begin{equation}
T\rightarrow 0:\;\;\;\;\;\; \Pi^{00}_{11}(0) \rightarrow 0\,\,.
\end{equation}
At zero temperature, 
static, homogeneous electric fields of colors 1, 2, or 3, are 
{\em not screened}. 

The other limit is when $T \rightarrow T_c$,
and $\phi \rightarrow 0$. Then, as $\int_0^\infty d \zeta / \cosh^2 \zeta
\equiv 1$,
\begin{equation}
T \rightarrow T_c: \;\;\;\;\;\; \Pi^{00}_{11}(0) \rightarrow
- 3\, m_g^2 \equiv {\Pi_0}^{00}(0) \,\,.
\end{equation}
As expected, $\Pi^{00}_{11}(0)$ approaches the value
in the normal phase, Eq.\ (\ref{Pi00HDLstatic}).

The interpretation of this result is the following.
From the explicit form of the Gell-Mann matrices it is clear that
gluons of adjoint colors 1, 2, and 3 ``see'' only quarks with fundamental
colors 1 and 2. However, at $T=0$, all these quarks are bound in Cooper pairs
to form a condensate of fundamental color (anti-)3, to which
these gluons are ``blind''. Hence, at $T=0$ the color superconductor
is transparent with respect to these color fields. There is nothing which
could screen these fields, thus there is no Debye mass for the gluons
of colors 1, 2, or 3.
Of course, this holds only in the limit $p_0 = 0$, $p\rightarrow 0$,
because only then are the gluons unable to resolve the individual
quarks (with colors that can be ``seen'') inside a Cooper pair.

When $T$ is nonzero, quasiparticles are thermally excited,
and screening sets in. As $T$ approaches $T_c$, the condensate
melts completely, and all quarks with the right colors to
screen gluon fields with colors 1, 2, and 3 are freed. Then,
the gluon self-energy approaches its value in the normal phase.
\\ ~~ \\
\noindent
\underline{(ii) $\mu =0,\, \nu =i$:}\hspace*{0.5cm} 
From Eq.\ (\ref{Pi110i}) it is clear that
\begin{equation} \label{Pi110istatic}
\Pi^{0i}_{11}(0,{\bf p}) \equiv 0\,\, .
\end{equation}
This is similar to the normal phase in the static limit,
Eq.\ (\ref{Pi0iHDLstatic}).
\\ ~~ \\
\noindent
\underline{(iii) $\mu =i,\,  \nu =j$:}\hspace*{0.5cm} 
As in the HDL limit, the magnetic components of the gluon self-energy
receive contributions not only from particle-particle and 
antiparticle-antiparticle, but also from particle-antiparticle excitations.
With Eq.\ (\ref{antiparticleapprox}) and $\int d\Omega \, \hat{k}^i\,
\hat{k}^j/(4 \pi) = \delta^{ij}/3$, one obtains from Eq.\ (\ref{Pi11ij})
\begin{eqnarray}
\Pi^{ij}_{11}(0) & \simeq & - \delta^{ij}\, \frac{g^2 \, N_f}{6 \pi^2}
\int_0^\infty dk\, k^2 \left\{ \frac{[\phi^+_{\bf k}]^2}{2 \,
[\epsilon^+_{\bf k}]^3}\,
\tanh \left( \frac{\epsilon^+_{\bf k}}{2T} \right) - 
\frac{dN(\epsilon^+_{\bf k})}{d \epsilon^+_{\bf k}}\,
\frac{\xi^2}{[\epsilon^+_{\bf k}]^2} \right. \nonumber \\
&   & \hspace*{3.1cm} + \left. \frac{4\, [1-N(\epsilon^+_{\bf k})]\, 
(1-n^+_{\bf k})}{\epsilon^+_{\bf k} + k + \mu}
- \frac{4\, N(\epsilon^+_{\bf k}) \, n^+_{\bf k}}{
\epsilon^+_{\bf k} - k - \mu} - \frac{2}{k} \right\} \,\, ,
\label{Pi11ijstatic}
\end{eqnarray}
where the last term was added to subtract the (UV-divergent) vacuum 
contribution, and where
$\phi^+_{\bf k} \equiv \phi^+(\epsilon^+_{\bf k},{\bf k})$.

At zero temperature, and after an integration by parts
($dn_{\bf k}^+/dk = - [\phi^+_{\bf k}]^2/2\, [\epsilon^+_{\bf k}]^3$),
\begin{equation}
\Pi^{ij}_{11}(0) \simeq  \delta^{ij}\, \frac{g^2 \, N_f}{6 \pi^2}
\int_0^\infty dk\, k \, \frac{4\, \epsilon_{\bf k}^+ }{
\epsilon_{\bf k}^+ + k + \mu} \, n_{\bf k}^+ \,
(1- n_{\bf k}^+)\,\, .
\end{equation}
The term $n_{\bf k}^+ \, (1- n_{\bf k}^+)$ is proportional to
$[\phi_{\bf k}^+]^2$. The momentum dependence of
the gap function is $\phi^+_{\bf k} = \phi_0 \, \sin (\bar{g}\, x_{\bf k})$
\cite{prlett2,prqcd},
where $\bar{g} = g/(3\sqrt{2} \pi)$ and $x_{\bf k} \simeq \ln [2b\mu/(
\epsilon_{\bf k}^+ + |\xi|)]$, with $\xi$ defined in Eq.\ (\ref{xi}) and
$b \equiv 256\, \pi^4[2/(N_f g^2)]^{5/2}$.
The gap function peaks at the Fermi surface, and is small far away
from the Fermi surface. Therefore, the region
$k \geq 2\, \mu$ can be neglected.

In the remaining integral over the region $0 \leq k \leq 2 \mu$,
take $k \simeq \mu$ in the slowly
varying factor $k/(\epsilon_{\bf k}^+ + k + \mu)$, and
change the integration variable to $\xi$:
\begin{equation} 
\Pi^{ij}_{11}(0) \simeq  \delta^{ij}\, \frac{g^2 \, N_f}{6 \pi^2}
\int_0^\mu \frac{d \xi}{\epsilon_{\bf k}^+} \, \left[\phi_{\bf k}^+
\right]^2 \,\,.
\end{equation}
Inserting the solution of the gap equation (including the momentum
dependence), and changing the integration variable
to $x = \ln [2b\mu/(\epsilon_{\bf k}^+ + \xi)]$, 
this integral can be solved analytically. However, it turns out that
this is unnecessary, if one only wants to know the
parametric dependence on the gap and
the QCD coupling constant in weak coupling, $g \ll 1$. 
One can simply neglect the
momentum dependence of the gap function, and approximate
$\phi_{\bf k}^+$ by its value at the Fermi surface, $\phi_0$, to
obtain
\begin{equation} \label{Pi11ijstaticT0}
\Pi^{ij}_{11}(0) \simeq  \delta^{ij}\,m_g^2\, \frac{\phi_0^2}{\mu^2} \,
 \ln \left( \frac{2\,\mu}{\phi_0}\right) \,\,.
\end{equation}
As $\phi_0 \sim \mu \, \exp(-c_{\rm QCD}/g)$,
$\Pi^{ij}_{11}$ is formally of order $\sim g \,\phi_0^2$.
To this order, I cannot exclude that there are cancellations
from other terms I have neglected (for instance the antiparticle
gaps). To leading order, the result (\ref{Pi11ijstaticT0})
is therefore consistent with $\Pi^{ij}_{11}(0) \simeq 0$.

Finally, as $T \rightarrow T_c$, an integration by parts shows
that the expression (\ref{Pi11ijstatic}) approaches the HDL limit,
Eq.\ (\ref{PiijHDLstatic}).

\subsection{Gluons with colors 4 and 6}

\noindent
\underline{(i) $\mu = \nu =0$:}\hspace*{0.5cm} 
For $p_0 = 0$, $p \rightarrow 0$, and with the approximations
(\ref{antiparticleapprox}), Eq.\ (\ref{Pi4400}) becomes
\begin{equation}
\Pi^{00}_{44} (0) \simeq - \frac{g^2\, N_f}{2 \pi^2}
\int_0^\infty dk \, k^2\, \left\{
   \frac{1-n^+_{\bf k}}{\epsilon^+_{\bf k} - \xi}
\, \left[N_F^+(k) - N( \epsilon^+_{\bf k}) \right]
+ \frac{n_{\bf k}^+}{\epsilon_{\bf k}^+ + \xi} 
\, \left[ 1 - N_F^+(k) - N( \epsilon^+_{\bf k}) \right] \right\} \,\,.
\label{Pi4400static}
\end{equation}
At $T = 0$, and restricting the $k$ integration to the range
$0 \leq k \leq 2 \mu$ (as before, the momentum dependence of the gap 
function suppresses any contribution from the region $k \geq 2 \mu$),
this can be transformed into
\begin{equation} 
\Pi^{00}_{44} (0) \simeq - 3 \, m_g ^2
\int_0^\mu \frac{d \xi}{\epsilon^+_{\bf k}} \, \left( 1 + \frac{\xi^2}{\mu^2}
\right) \, \frac{ \epsilon^+_{\bf k} - \xi}{\epsilon^+_{\bf k} + \xi} \,\, .
\end{equation}
Neglecting the momentum dependence of the gap function, the
remaining integral can be done introducing the variable
\begin{equation}
y \equiv \ln \left( \frac{ \epsilon_{\bf k}^+ + \xi}{\phi_0} \right)\,\,.
\end{equation}
To leading order, the result is
\begin{equation} \label{Pi4400staticT0}
\Pi^{00}_{44} (0) \simeq - \frac{3}{2} \, m_g ^2 \,\, .
\end{equation}
The Debye mass is reduced by a factor 2 as compared to the value
in the normal phase.

The limit $T \rightarrow T_c$ cannot be studied
with Eq.\ (\ref{Pi4400static}), and one has to go back
to Eq.\ (\ref{Pi4400}). It is obvious that one will reproduce
the HDL result (\ref{Pi00HDLstatic}).
\\ ~~ \\
\noindent
\underline{(ii) $\mu = 0,\, \nu =i$:}\hspace*{0.5cm} 
With Eq.\ (\ref{Pi440i}), and the same approximations as before,
one obtains
\begin{eqnarray}
\Pi^{0i}_{44} (0) & \simeq & - \frac{g^2\, N_f}{2 \pi^2}
\int_0^\infty dk \, k^2\, \int \frac{d \Omega}{4 \pi} \, \hat{k}^i\,
   \left\{
   \frac{1-n^+_{\bf k}}{\epsilon^+_{\bf k} - \xi}
\, \left[N_F^+(k) - N( \epsilon^+_{\bf k}) \right]
+ \frac{n_{\bf k}^+}{\epsilon_{\bf k}^+ + \xi} 
\, \left[ 1 - N_F^+(k) - N( \epsilon^+_{\bf k}) \right] \right\} 
\nonumber \\
& \equiv & 0\,\,,
\label{Pi440istatic}
\end{eqnarray}
by symmetry.
\\ ~~ \\
\noindent
\underline{(iii) $\mu = i,\, \nu =j$:}\hspace*{0.5cm} 
From Eq.\ (\ref{Pi44ij}) one derives under the same approximations
\begin{eqnarray}
\Pi^{ij}_{44}(0) & \simeq & - \delta^{ij}\, \frac{g^2\, N_f}{6 \pi^2}
\int_0^\infty dk \, k^2 \left\{ 
   \frac{1-n^+_{\bf k}}{\epsilon^+_{\bf k} - \xi}
\, \left[N_F^+(k) - N( \epsilon^+_{\bf k}) \right]
+ \frac{n_{\bf k}^+}{\epsilon_{\bf k}^+ + \xi} 
\, \left[ 1 - N_F^+(k) - N( \epsilon^+_{\bf k}) \right] \right.
\nonumber \\
&     & + \left. \frac{1}{k}\, \left[ 1-N_F^+(k)\right] 
+ 2\, \frac{1-n^+_{\bf k}}{\epsilon^+_{\bf k} + k + \mu}\,
\left[1 - N( \epsilon^+_{\bf k}) \right] 
- 2\, \frac{n^+_{\bf k}}{\epsilon^+_{\bf k} - k - \mu}\,
N(\epsilon^+_{\bf k})  - \frac{2}{k} \right\} \,\,,
\label{Pi44ijstatic}
\end{eqnarray}
where the last term is a vacuum subtraction.

At $T=0$, the integral over the first two terms in the integrand has
already been computed for $\Pi^{00}_{44}(0)$, with the result
(\ref{Pi4400staticT0}). This is cancelled by a part of the vacuum subtraction.
The remainder is
\begin{equation} \label{121}
\Pi^{ij}_{44}(0)  \simeq  \delta^{ij}\, \frac{g^2\, N_f}{6 \pi^2}
\int_0^\infty dk \, \frac{k}{\epsilon^+_{\bf k}}\,
\frac{\mu\, (\epsilon^+_{\bf k} - \xi) + \left[\phi^+_{\bf k} 
\right]^2}{\epsilon^+_{\bf k} + \xi + 2\mu}\,\,.
\end{equation}
Because the momentum dependence of the gap function suppresses
the contribution from momenta far from the Fermi surface,
the integral can be restricted to the region $0 \leq k \leq 2\, \mu$.
To leading order, one may neglect $\left[ \phi^+_{\bf k} \right]^2$
in the numerator. [The respective contribution is of order $\phi_0^2
\ln (2 \mu/\phi_0)$.]
Then, introduce the integration variable $z = \epsilon_{\bf k}^+ - k + \mu$.
Neglecting the momentum dependence of the gap function, as well as
terms of order $[\phi^+_{\bf k}]^2$, one obtains
\begin{equation}
\Pi^{ij}_{44}(0)  \simeq  \delta^{ij}\, \frac{g^2\, N_f}{12 \pi^2}\,
\mu \, \int_0^{2 \mu} dz \, \left( 1 - \frac{z}{2\mu} \right)
= \delta^{ij}\, \frac{m_g^2}{2} \,\,.
\end{equation}
The limit $T \rightarrow T_c$ is not well-defined for 
Eq.\ (\ref{Pi44ijstatic}); using Eq.\ (\ref{Pi44ij})
it is, however, straightforward to show that 
$\Pi^{ij}_{44}(0) \rightarrow {\Pi_0}^{ij}(0)$, as expected.

\subsection{Gluons with color 5 and 7}

In the limit $p_0 = 0$, $p \rightarrow 0$, {\it i.e.},
${\bf k}_2 \rightarrow {\bf k}_1$, it is obvious
from comparing Eqs.\ (\ref{Pi44b}) and (\ref{Pi55b}) that
\begin{equation}
\Pi^{\mu \nu}_{44}(0) \equiv \Pi^{\mu \nu}_{55}(0)\,\, ,
\end{equation}
hence, the results from the previous subsection can be carried over.

\subsection{Gluons with color 8}

\noindent
\underline{(i) $\mu = \nu =0$:}\hspace*{0.5cm} 
From Eq.\ (\ref{Pi8800}) one obtains with the approximations
(\ref{antiparticleapprox})
\begin{equation}
\tilde{\Pi}^{00}(0) \simeq \frac{g^2 \, N_f}{2 \pi^2} \int_0^\infty 
dk \, k^2 \, \left\{
\frac{d n^+_{\bf k}}{d k} \, \left[ 1- 2 N( \epsilon^+_{\bf k}) \right]
+ \frac{d N(\epsilon^+_{\bf k})}{dk} \, \left( 1- 2 n^+_{\bf k}\right) 
\right\}\,\, .
\end{equation}
The integrand is vanishingly small except close to the Fermi
surface. One can therefore restrict the
$k$ integration to the range $0 \leq k \leq 2\mu$.
Then, introducing $\xi$ as integration variable and using the
symmetry of the integrand around $\xi = 0$,
\begin{equation}
\tilde{\Pi}^{00}(0) \simeq -3\, m_g^2 \int_0^{\mu}
d\xi\, \frac{d}{d\xi} \left[ \frac{ \xi}{\epsilon^+_{\bf k}} \,
\tanh \left( \frac{ \epsilon^+_{\bf k}}{2T} \right) \right] \,\, ,
\end{equation}
where higher order terms ($ \sim \xi^2/\mu^2$)
in the integrand have been neglected.
The remaining integral is unity (remember that $ \mu \gg T$), and the
final result is
\begin{equation}
\tilde{\Pi}^{00}(0) \simeq -3\, m_g^2 \,\,.
\end{equation}
Note that this result is independent of the temperature.
One concludes that 
\begin{equation}
\Pi^{00}_{88}(0) \equiv \frac{2}{3}\, {\Pi_0}^{00}(0) + 
\frac{1}{3}\, \tilde{\Pi}^{00}(0) \equiv - 3\, m_g^2
\end{equation}
does not change with temperature in the superconducting phase; 
it always has the same value as in the normal phase.
\\ ~~ \\
\noindent
\underline{(ii) $\mu =0,\,  \nu =i$:}\hspace*{0.5cm} 
On account of Eqs.\ (\ref{Pi880i}) and (\ref{Pi110istatic}),
\begin{equation}
\tilde{\Pi}^{0i}(0,{\bf p}) \simeq 0 \,\, .
\end{equation}
Consequently, also $\Pi^{0i}_{88}(0) \simeq 0$.
\\ ~~ \\
\noindent
\underline{(iii) $\mu = i,\, \nu =j$:}\hspace*{0.5cm} 
For $\tilde{\Pi}^{ij}(0)$ one derives from Eq.\ (\ref{Pi88ij}) with
the standard approximations
\begin{equation}
\tilde{\Pi}^{ij}(0) \simeq  - \, \delta^{ij}\, \frac{g^2 \, N_f}{6 \pi^2}
\int_0^\infty dk\, k^2 \left\{ - \frac{dN(\epsilon^+_{\bf k})}{d
\epsilon^+_{\bf k}} + \frac{4\, [1-N(\epsilon^+_{\bf k})]\, 
(1-n^+_{\bf k})}{\epsilon^+_{\bf k} + k + \mu}
- \frac{4\, N(\epsilon^+_{\bf k}) \, n^+_{\bf k}}{
\epsilon^+_{\bf k} - k - \mu} - \frac{2}{k} \right\} \,\, ,
\label{Pi88ijstatic}
\end{equation}
where the last term is a vacuum subtraction.

At $T=0$, Eq.\ (\ref{Pi88ijstatic}) becomes twice the
integral in Eq.\ (\ref{121}), hence
\begin{equation}
\tilde{\Pi}^{ij}(0) \simeq \delta^{ij}\, m_g^2\,\, .
\end{equation}
As a consequence,
\begin{equation}
\Pi^{ij}_{88} (0) \simeq \delta^{ij}\, \frac{m_g^2}{3}\,\, .
\end{equation}

As $T \rightarrow T_c$, an integration by parts shows that
$\tilde{\Pi}^{ij}(0) \rightarrow 0$, as it should be.
Consequently, also $\Pi^{ij}_{88}(0) \rightarrow 0$.

This concludes the discussion of Debye screening and the Meissner
effect. In the next section, it will be demonstrated that
for momenta $p \gg \phi_0$, {\it i.e.}, when the gluon momentum is
large enough to resolve the quarks in a Cooper pair,
the gluon self-energy approaches the value in the normal phase.

\section{Nonzero gluon momentum}\label{vi}

In this section, the gluon self-energy will be computed in the
static limit, but for gluon momenta $\phi_0 \ll p \ll \mu$.
In the condensed matter literature, this limit is known
as the Pippard limit \cite{fetter}. The actual calculation
follows closely that for ordinary superconductors (see for instance
\cite{fetter}). It will be convenient to consider
the difference between the self energies in the superconducting
and normal phases,
\begin{equation}
\delta \Pi \equiv \Pi - {\Pi_0}\,\, .
\end{equation}

For large gluon momenta, effects from the pairing of quarks have
to vanish, as the gluon wave length is short enough to resolve
individual quarks in a Cooper pair. Consequently, the Debye mass
for gluons of color 1, 2, and 3 can no longer vanish, but must
approach the value in the normal phase. Simultaneously, for
gluons of color 8 the Meissner effect has to vanish. These are the
two cases studied in this section.

Of course, also the electric and magnetic masses of gluons with colors 
4, 5, 6, and 7 have to approach their values in the normal phase.
I was, however, not able to derive
simple analytical expressions for the self-energy of these gluons
in the limit $\phi_0 \gg p \gg \mu$.
An explicit numerical study will be deferred to the future.

First note that for $p \ll \mu$, $k \sim \mu$,
\begin{equation}
k_{1,2} \simeq k \pm \frac{\hat{\bf k} \cdot {\bf p}}{2} \,\, .
\end{equation}
This then leads to the same expressions (\ref{43a})
-- (\ref{43c}) for the spin traces as in the HDL limit. 
As in the previous section, 
quasi-antiparticles will be treated as real antiparticles, cf.\ Eq.\ 
(\ref{antiparticleapprox}). Furthermore, for the sake of notational
convenience, let us introduce
\begin{equation}
\xi_\pm  \equiv  \xi \pm \frac{\hat{\bf k} \cdot {\bf p}}{2} \;\;\;\; ,
\;\;\;\; \epsilon_\pm \equiv \epsilon^+_{{\bf k}_{1,2}} \;\;\;\; , 
\;\;\;\; \phi_\pm \equiv \phi^+(\epsilon_\pm) \;\;\;\; , \;\;\;\;
n_\pm  \equiv  n^+_{{\bf k}_{1,2}} \;\;\;\; , \;\;\;\;
N_\pm \equiv N(\epsilon_\pm)\,\,.
\end{equation}

\subsection{Electric gluons of color 1, 2, and 3}\label{via}

Writing $N_\pm = [1-\tanh(\epsilon_\pm/2T)]/2$, the self-energy
of electric gluons of colors 1, 2, and 3 is from Eq.\ (\ref{Pi1100})
\begin{eqnarray}
\Pi^{00}_{11} (0, {\bf p}) & \simeq & - \frac{g^2 \, N_f}{2}
\int \frac{d^3 {\bf k}}{(2 \pi)^3} \left\{
\frac{1}{\epsilon_+ + \epsilon_-} \,\left[
\tanh \left( \frac{\epsilon_+}{2T} \right) 
+ \tanh \left(\frac{\epsilon_-}{2T} \right) \right]\,
\frac{1}{2} \left(1- \frac{ \xi_+ \, \xi_- + \phi_+\, \phi_-}{
\epsilon_+ \, \epsilon_-} \right) \right. \nonumber \\
&    & \hspace*{2.55cm} + \left. \frac{1}{\epsilon_+ - \epsilon_-} \, \left[
\tanh \left( \frac{\epsilon_+}{2T} \right) 
- \tanh \left(\frac{\epsilon_-}{2T} \right) \right]\,
\frac{1}{2} \left(1+ \frac{ \xi_+ \, \xi_- + \phi_+\, \phi_-}{
\epsilon_+ \, \epsilon_-}\right) \right\} \,\, , \label{Pi1100p}
\end{eqnarray}
where terms of order $p^2/k^2$ have been neglected.
The self-energy in the normal phase can be obtained
either from Eq.\ (\ref{Pi2}), for $p_0 = 0$ and with the
approximations (\ref{antiparticleapprox}), or directly from
Eq.\ (\ref{Pi1100p}) in the limit $\phi_\pm \rightarrow 0$:
\begin{equation} \label{Pi00HDLp}
{\Pi_0}^{00}(0, {\bf p}) \simeq - \frac{g^2 \, N_f}{2}
\int \frac{d^3 {\bf k}}{(2 \pi)^3} 
\frac{1}{\xi_+ - \xi_-} \,\left[
\tanh \left( \frac{\xi_+}{2T} \right) 
- \tanh \left(\frac{\xi_-}{2T} \right) \right]\,\, .
\end{equation}
Now consider the difference $\delta \Pi^{00}_{11}(0,{\bf p})$ 
between (\ref{Pi1100p}) and (\ref{Pi00HDLp}).
As the main contribution to the integral over ${\bf k}$ comes
from the region around the Fermi surface, it is admissible to
neglect the momentum dependence of the gap function,
$\phi_+ \simeq \phi_- \equiv \phi$. Then one rearranges the integrand
to separate terms of the form
\begin{equation} \label{odd}
\frac{1}{\xi_+ - \xi_-} \,\left[ \frac{\xi_\pm}{\epsilon_\pm}\,
\tanh \left( \frac{\epsilon_\pm}{2T} \right) 
- \tanh \left(\frac{\xi_\pm}{2T} \right) \right] \,\, .
\end{equation}
As argued in \cite{fetter}, these terms vanish by symmetry
when integrating over $\xi$. (A careful analysis shows
that this is correct to leading order in $\phi/p$.)
The result is
\begin{equation}
\delta \Pi^{00}_{11}(0,{\bf p}) \simeq - \frac{g^2 \, N_f}{2}
\int \frac{d^3 {\bf k}}{(2 \pi)^3} 
\frac{\phi^2}{\xi\, \hat{\bf k} \cdot {\bf p}} \,
\left[ \frac{1}{\epsilon_+}\,\tanh \left( \frac{\epsilon_+}{2T} \right) 
- \frac{1}{\epsilon_-}\, \tanh \left(\frac{\epsilon_-}{2T} \right) \right]
\,\, .
\end{equation}
As the integrand peaks at the Fermi surface, $\xi \simeq 0$, and for
$\hat{\bf k} \cdot {\bf p} \simeq 0$, one can approximate the
hyperbolic tangens by $\tanh (\epsilon_\pm/2T) \sim \tanh (\phi/2T)$,
and obtains to leading order 
\begin{equation} \label{thesame}
\delta \Pi^{00}_{11}(0,{\bf p}) \simeq - 3 \, m_g^2\,
\frac{\phi}{p} \, \tanh \left( \frac{\phi}{2T} \right)
\int_0^{\mu/\phi} \frac{d x}{x} \int_0^{p/2\phi} \frac{dy}{y}
\left( \frac{1}{\sqrt{(x+y)^2+1}} - \frac{1}{\sqrt{(x-y)^2+1}} \right)\,\, ,
\end{equation}
where $x \equiv \xi/\phi$, $y \equiv \hat{\bf k} \cdot {\bf p}/(2 \phi)$.
The $y$ integral can be done exactly.
In the limit $\mu \gg p \gg \phi$,
\begin{equation}
\delta \Pi^{00}_{11}(0,{\bf p}) \simeq 3 \, m_g^2\,
\frac{\phi}{p} \, \tanh \left( \frac{\phi}{2T} \right)
 \int_0^{\infty} d u \, \frac{2 u}{\sinh u} 
\equiv 3\, m_g^2 \, \frac{\pi^2}{2}\, 
\frac{\phi}{p} \, \tanh \left( \frac{\phi}{2T} \right)\,\, .
\end{equation}
The self-energy in the normal phase is approximately constant
for momenta $p \ll \mu$, such that
\begin{equation} \label{Pi1100pfinal}
\Pi^{00}_{11}(0,{\bf p}) \simeq - 3\, m_g^2 \left[1-
\frac{\pi^2}{2}\, \frac{\phi}{p} \, \tanh \left( \frac{\phi}{2T} \right)
\right]\,\,.
\end{equation}
This shows that the absolute value of the
self-energy in the superconducting phase is reduced as compared
to the normal phase. For increasing $p/\phi$, the correction 
becomes smaller,
such that electric fields for adjoint colors 1, 2, and 3 are
screened over an only slightly longer distance than in the normal
phase. In this case, the gluons ``see'' the individual fundamental
color charges inside the Cooper pairs.

For decreasing $p/\phi$, however, the correction becomes larger.
This is in agreement with the results of
Sec.\ \ref{v}, where the self-energy of gluons with colors 1, 2, and 3
was found to vanish in the limit $p \rightarrow 0$, {\it i.e.},
when the gluon momentum is too small to resolve individual quarks 
inside a Cooper pair. Although strictly valid only for $p \gg \phi$,
by extrapolating Eq.\ (\ref{Pi1100pfinal}) to $p \sim \phi$ one would conclude
that, at $T=0$, this happens once $p$ is smaller than $\simeq 5 \, \phi_0$. 

\subsection{Magnetic gluons of color 8}\label{vib}

For magnetic gluons, one derives from Eq.\ (\ref{Pi88ij})
\begin{eqnarray}
\tilde{\Pi}^{ij} (0, {\bf p}) & \simeq & - \frac{g^2 \, N_f}{2}
\int \frac{d^3 {\bf k}}{(2 \pi)^3} \left( \hat{k}^i \, \hat{k}^j
\left\{ \frac{1}{\xi_+ - \xi_-} \,\left[
\frac{\xi_+}{\epsilon_+}\, \tanh \left( \frac{\epsilon_+}{2T} \right) 
- \frac{\xi_-}{\epsilon_-}\, 
\tanh \left(\frac{\epsilon_-}{2T} \right) \right] \right. \right.
\nonumber \\
&    & \hspace*{3.85cm} + \left.
\frac{\phi^2}{\xi \, \hat{\bf k} \cdot {\bf p}}\,
\left[\frac{1}{\epsilon_+}\, \tanh \left( \frac{\epsilon_+}{2T} \right) 
- \frac{1}{\epsilon_-}\, \tanh \left( \frac{\epsilon_-}{2T} \right) \right]
\right\}  \nonumber \\
&    & \hspace*{0.7cm} + 
\left(\delta^{ij} - \hat{k}^i\, \hat{k}^j \right) \,
\frac{1}{2k} \, \left\{ 2 + \frac{\xi_+}{\epsilon_+}\,
\tanh \left( \frac{\epsilon_+}{2T} \right) 
+ \frac{\xi_-}{\epsilon_-}\,
\tanh \left(\frac{\epsilon_-}{2T} \right) \right. \nonumber \\
&    & \hspace*{3.55cm} - \left. \left.  
\frac{\phi^2}{2\mu} \left[\frac{1}{\epsilon_+}\,
\tanh \left( \frac{\epsilon_+}{2T} \right) 
+ \frac{1}{\epsilon_-}\,
\tanh \left(\frac{\epsilon_-}{2T} \right) \right] \right\} \right)
\,\, . \label{Pi88ijp}
\end{eqnarray}
Here, the momentum dependence of the gap function was neglected,
$\phi_\pm \simeq \phi$. Moreover, in denominators which contain
terms $\sim \mu^2$, $\epsilon_\pm^2$ was approximated by 
$\xi_\pm^2$.

In the normal phase, the corresponding expression reads
\begin{eqnarray}
{\Pi_0}^{ij} (0, {\bf p}) & \simeq & - \frac{g^2 \, N_f}{2}
\int \frac{d^3 {\bf k}}{(2 \pi)^3} \left\{ \hat{k}^i \, \hat{k}^j
 \frac{1}{\xi_+ - \xi_-} \,\left[
\tanh \left( \frac{\xi_+}{2T} \right) - 
\tanh \left(\frac{\xi_-}{2T} \right) \right] \right.
\nonumber \\
&    & \hspace*{1.7cm} + \left.
\left(\delta^{ij} - \hat{k}^i\, \hat{k}^j \right) \,
\frac{1}{2k} \, \left[ 2 + \tanh \left( \frac{\xi_+}{2T} \right) 
+ \tanh \left(\frac{\xi_-}{2T} \right) \right] \right\}
\,\, . \label{PiijHDLp}
\end{eqnarray}
In the difference $\delta \tilde{\Pi}^{ij}$, there are again terms
like (\ref{odd}), which vanish by symmetry arguments. 
There is also a term $\sim \phi^2/(4 \mu k)$ which is
of higher order and thus can be neglected. The remainder
can be written as
\begin{eqnarray}
\lefteqn{\delta \tilde{\Pi}^{ij} (0, {\bf p}) \simeq -
 3\, m_g^2 \, \frac{\phi}{p} } \nonumber \\
&  \times & \int_0^{\mu/\phi}
\frac{dx}{x} \int_0^{p/2 \phi} \frac{dy}{y}
\tanh \left( \frac{ \phi \, \sqrt{y^2 + 1}}{2T} \right)
\left( \frac{1}{\sqrt{(x+y)^2 + 1}} - \frac{1}{\sqrt{(x-y)^2 + 1}} 
\right) \int_0^{2 \pi} \frac{d\varphi}{2 \pi} \, \hat{k}^i \, \hat{k}^j \,\,.
\label{Pi88ijpfinal}
\end{eqnarray}
As before, $x \equiv \xi/ \phi$, $y \equiv \hat{\bf k} \cdot {\bf p}/2 \phi$.
Since the $x$ integral is dominated by the region around the Fermi surface,
$x \simeq 0$, I have set $x = 0$ in the argument of the hyperbolic
tangens.

For $i \neq j$, the integration over the polar angle $\varphi$ vanishes,
thus $\delta \tilde{\Pi}^{ij}$ is diagonal. However, not all
diagonal elements are equal. Let ${\bf p} = (0,0,p)$.
Then $\hat{k}_x^2 = [1-(2 \phi \,y/p)^2] \cos^2 \varphi$,
$\hat{k}_y^2 = [1-(2 \phi\, y/p)^2] \sin^2 \varphi$, and
the transverse components of $\delta \tilde{\Pi}^{ij}$ are
\begin{equation}
\delta \tilde{\Pi}^{xx} (0, {\bf p})
\equiv \delta \tilde{\Pi}^{yy}(0,{\bf p}) \simeq 
 m_g^2 \, \frac{3 \pi^2}{4}\, \frac{\phi}{p} 
\tanh \left( \frac{ \phi}{2T} \right)\,\,.
\label{Pi88xxpfinal}
\end{equation}
To obtain this result, I have used the fact that the $y$ integration
is dominated by the region $y\simeq 0$, and consequently
have set $y=0$ in the hyperbolic tangens as well as in
$\hat{k}^2_{x,y}$. The remaining integral is then the same as
in Eq.\ (\ref{thesame}).

The longitudinal component can be shown to be of higher order
in $\phi/p$, such that to leading order,
\begin{equation}
\delta \tilde{\Pi}^{zz}(0, {\bf p}) \simeq 0 \,\,.
\label{Pi88zzpfinal}
\end{equation}
This result is not unexpected: the self-energy for
gluons in the normal phase is transverse,
${\Pi_0}^{ij}(0,{\bf p}) \simeq(\delta^{ij} - \hat{p}^i\, \hat{p}^j)\,
p^2 \, m_g^2/(12\, \mu^2)$. [Note that this expressions is of order
$g^2 p^2 \ll g^2 \mu^2$, and thus not in contradiction to the HDL
result (\ref{PiijHDLstatic}).]
Equations (\ref{Pi88xxpfinal}) and (\ref{Pi88zzpfinal}) now
combine to give a transverse self-energy for the eighth gluon, too,
\begin{equation}
\Pi^{ij}_{88} (0, {\bf p}) \simeq
\left(\delta^{ij} - \hat{p}^i\, \hat{p}^j \right)\, m_g^2
\, \left[ \frac{p^2}{12\, \mu^2}
+ \frac{\pi^2}{4}\, \frac{\phi}{p}\, \tanh \left( \frac{\phi}{2T}
\right) \right]\,\,. 
\end{equation}

\section{Summary, conclusions, and outlook} \label{vii}

In color-superconducting quark matter with $N_f =2$ degenerate
quark flavors, the condensate can be oriented 
in (anti-)3 direction in fundamental color space by
means of a global color rotation. Then,
only quarks with fundamental colors 1 and 2 form Cooper pairs, while
quarks of the third fundamental color remain unpaired, and act
as a background to neutralize the color-charged condensate.
Since the unpaired quarks carry the same 
color charge, two of them are
in the (repulsive) sextet representation of $SU(3)_c$. Consequently,
they do not form Cooper pairs and the system is stable.

The condensate breaks the $SU(3)_c$ color symmetry to
$SU(2)_c$. With the above color choice, the generators of the
unbroken $SU(2)_c$ subgroup are the $SU(3)_c$ generators
$T^1,\, T^2$, and $T^3$, with $T^a = \lambda^a/2$ and the
standard convention for the Gell-Mann matrices $\lambda^a$.
The gluons corresponding to the remaining generators $T^4$ through 
$T^8$ all receive a mass via the Anderson--Higgs mechanism.

What are the expected values for these masses?
The effective Lagrangian for the low-energy excitations
of the condensate fields minimally coupled to gauge fields has the
kinetic term \cite{rp}
\begin{equation}
{\cal L}_{\rm eff}^{\rm kin} = \alpha_{\rm e} \,  \left(
D_0 \Phi \right)^\dagger D^0 \Phi
+ \alpha_{\rm m} \, \left( D_i \Phi \right)^\dagger D^i \Phi \,\, .
\end{equation}
The presence of a heat and particle bath at nonzero $T$ and/or $\mu$ breaks
Lorentz invariance, so that the coefficient $\alpha_{\rm e}$
of the part containing the time derivatives can in principle be 
different from the one of the part containing the spatial derivatives,
$\alpha_{\rm m}$.

For a two-flavor color-superconductor, $\Phi$ is
a $SU(3)_c$ (anti-)triplet, $\Phi \equiv (\Phi_1, \Phi_2, \Phi_3)^T$ 
\cite{prlett}. Consequently, the
covariant derivative is $D_\mu = \partial_\mu
- ig A_\mu^a T^a$, with the generators $T^a$ being in the fundamental
representation.
If $\Phi$ attains a non-vanishing
expectation value $\langle \Phi \rangle = (0,0,\phi_0)^T$, 
$\phi_0 \in {\bf R}$, this generates a mass term for the gluon fields 
of the form
\begin{eqnarray} \label{kineff}
{\cal L}_1^{\rm M}  & = & g^2 \, \phi_0^2 \left(\alpha_{\rm e} \,
 A_0^a \, A^0_b + \alpha_{\rm m}\, A_i^a \, A^i_a \right)\; 
\delta_{3i}\, T^a_{ij}\,  T^b_{jk}\, \delta_{k3} \nonumber \\
& \equiv &  g^2\, \phi_0^2  \, \left[
\frac{1}{4} \sum_{a=4}^7  \left( \alpha_{\rm e}\, A_0^a \, A^0_a
+ \alpha_{\rm m} \, A_i^a\, A^i_a \right)
+ \frac{1}{3} \, \left( \alpha_{\rm e}\, A_0^8 \, A^0_8
+ \alpha_{\rm m} \, A_i^8 \, A^i_8 \right) \right]
\,\,.
\end{eqnarray}
The expected electric and magnetic gluon masses are
\begin{equation} \label{individualmasses}
M_{\rm e,m}^1 = M_{\rm e,m}^2 = M_{\rm e,m}^3 = 0 \;\;\;\; , \;\;\;\;
M_{\rm e,m}^4 = M_{\rm e,m}^5 = M_{\rm e,m}^6 = M_{\rm e,m}^7 =
\sqrt{\frac{\alpha_{\rm e,m}}{2}} \, g \, \phi_0  \;\;\;\; , \;\;\;\;
M_{\rm e}^8 = \sqrt{\frac{2 \,\alpha_{\rm e,m}}{3} }\, g \, \phi_0 \,\, ,
\end{equation}
such that the ratio 
\begin{equation}\label{expectedmass}
R_{\rm e,m} \equiv \left( \frac{M_{\rm e,m}^8}{ M_{\rm e,m}^4}\right)^2 = 4/3
\,\,.
\end{equation}

In this work, the gluon self-energy in a
$N_f=2$ color superconductor has been derived.
Due to the pattern of symmetry breaking, one has to study the
individual gluon colors separately.
The central result are equations (\ref{Pi11all}) -- (\ref{Pi88all}). 
Various limits of these expressions are of interest.
Here, the self-energy was computed in the static, homogeneous limit,
$p_0 = 0$, $p \rightarrow 0$, which yields the Debye mass for
electric and the Meissner mass for magnetic gluons. The main
results are summarized in Table \ref{Table1}.

For the three gluons of the unbroken $SU(2)_c$ subgroup
(gluon colors 1, 2, and 3), the Debye mass as well as the Meissner 
mass vanish. While this is in agreement with (\ref{individualmasses}),
it is at first physically unclear, and therefore quite surprising, 
why gluon fields with colors 1, 2, and 3 are not screened.
To explain this, I argued as follows.
Gluons with adjoint colors 1, 2, and 3 couple to fundamental colors 1 and 2.
At $T=0$, however, all quarks with these color charges are bound
in Cooper pairs which have fundamental color (anti-)3. Thus, these gluons
cannot ``see'' the quark charges, and hence are unscreened.
At nonzero $T$, quasiparticles are thermally excited.
They have the ``right'' fundamental color (1 and 2)
to screen gluon fields with adjoint colors 1, 2, and 3, and
consequently lead to screening and a nonzero Debye mass. 
At $T= T_c$, when the condensate melts, the
Debye mass assumes its standard value in the normal phase.

Of course, at $T=0$ the gluon self-energy vanishes only
in the zero-energy, zero-momentum limit,
since then the gluon field cannot resolve individual quarks inside
the Cooper pair. For large gluon momentum $p \gg \phi_0$, electric gluon
fields are screened; the self-energy is the same as
in the normal phase, up to a correction of
order $\sim m_g^2 \, \phi_0/p$, as computed in Sec.\ \ref{via}.

The gluons corresponding to the broken generators of $SU(3)_c$ all
attain a mass through the Anderson--Higgs mechanism. 
While the Debye mass for electric gluons of color 8 is the same
as in the normal phase, the Debye mass squared for colors 4 through 7 is only
half as large. As $T$ approaches $T_c$, however, the melting of the
condensate leads to an increase of the Debye mass to its standard
value. At zero temperature, the ratio of the Debye masses squared of 
gluon color 8 and 4 is 
$R_{\rm e} \equiv \Pi^{00}_{88}(0)/\Pi^{00}_{44}(0) = 2$.

The Meissner mass squared for gluons of color 8 is 
$1/3$ of the gluon mass squared, $m_g^2$, 
while that for gluons of colors 4 through
7 is $1/2$ of the gluon mass squared. The Meissner effect vanishes as
$T$ approaches $T_c$, or when the gluon momentum $p \gg \phi_0$,
as computed in Sec.\ \ref{vib}. The ratio of the Meissner masses
squared of gluon color 4 and 8 is
$R_{\rm m} \equiv \Pi^{ii}_{88}(0)/\Pi^{ii}_{44}(0) = 2/3$.

Both $R_{\rm e}$ and $R_{\rm m}$ differ from the expectation 
(\ref{expectedmass}). What is the origin of this discrepancy?
The kinetic term (\ref{kineff}) is not the only possible invariant
in an effective Lagrangian, where the condensate fields are
minimally coupled to the gauge fields. Another possibility is
the term \cite{shovkovy}
\begin{equation} \label{kineff2}
{\cal L}_{\rm eff}' = \beta_{\rm e} \, 
\left( \Phi^\dagger D_0 \Phi \right)^\dagger
\Phi^\dagger D^0 \Phi + \beta_{\rm m} \, 
\left( \Phi^\dagger D_i \Phi \right)^\dagger
\Phi^\dagger D^i \Phi \,\, ,
\end{equation}
which has mass dimension six [consequently, $\beta_{\rm e,m}$ have dimension 
(mass)$^{-2}$]. Note that in the nonlinear version of the effective
theory \cite{sonstephanov}, where the modulus of $\Phi$ does not change,
only the phase, this term is identical to the standard kinetic term
(\ref{kineff}).

\begin{table}[h]
\caption{Results for the Debye and Meissner masses in a
two-flavor color superconductor.}
\begin{tabular}{||c||cccc|cccc||}
            &    &       &        &      &       &      &      & \\ 
gluon color &                & $-\Pi^{00}_{aa}(0)$ &     &        &
                             & $\Pi^{ii}_{aa}(0)$  &      &       \\ 
            &    &        &       &      &        &     &      &  \\ 
 $a$      &    $T=0$   &  & $T \geq T_c$   & &  $T=0$ &  & $T \geq T_c $ &\\ 
            &    &        &       &      &        &     &      &  \\ 
    \hline \hline 
            &    &        &       &      &        &     &      &  \\ 
1 -- 3    &      0     &  &  $3\, m_g^2$   & &    0   &  &   0          &   \\ 
            &    &        &       &      &        &     &      &  \\  
     \hline
            &    &        &       &      &        &     &      &  \\ 
4 -- 7    & $\frac{3}{2}\ m_g^2$ & & $3\, m_g^2$ & & $\frac{1}{2}\,m_g^2$ & &
    0  & \\ 
            &    &        &       &      &        &     &      &  \\  
     \hline
            &    &        &       &      &        &     &      &  \\  
8         & $3\, m_g^2 $  & & $3\, m_g^2$   & & $\frac{1}{3}\,m_g^2$ & & 0 & \\
            &    &        &       &      &        &     &      &    
\end{tabular}
\label{Table1}
\end{table}

Upon condensation, $\langle \Phi \rangle = (0,0,\phi_0)^T$, the
term (\ref{kineff2}) contributes to the mass of the eighth gluon,
\begin{equation}
{\cal L}_2^{\rm M} =  g^2 \, \phi_0^4 \, \frac{1}{3}\, 
\left( \beta_{\rm e}\, A_0^8 \, A^0_8  + \beta_{\rm m}\,
A_i^8 \, A^i_8 \right)\,\, .
\end{equation}
With this term, one reproduces the zero-temperature magnetic masses 
given in Table \ref{Table1} with the choice 
\begin{equation} \label{magnetic}
\alpha_{\rm m} \equiv \frac{m_g^2}{g^2 \, \phi_0^2} = \frac{N_f}{6\pi^2}\,
\frac{\mu^2}{\phi_0^2} \;\;\;\; , \;\;\;\;
\beta_{\rm m} \equiv - \frac{1}{2}\, \frac{m_g^2}{g^2 \, \phi_0^4}
= - \frac{N_f}{12\pi^2}\, \frac{\mu^2}{\phi_0^4} \,\, .
\end{equation}
Note that the prefactor of the kinetic term (\ref{kineff})
has the $1/\phi_0^2$ behavior typical for effective theories
of superconductivity \cite{bailinlove,fetter,rp}.
To reproduce the electric masses,
the coefficients $\alpha_{\rm e}$ and $\beta_{\rm e}$ have to be chosen as
\begin{equation} \label{electric}
\alpha_{\rm e} \equiv 3\, \alpha_{\rm m} \;\;\;\; , \;\;\;\; 
\beta_{\rm e} = - 3\, \beta_{\rm m}\,\,.
\end{equation}

The expressions (\ref{magnetic}) and (\ref{electric}) fix the
prefactors of the kinetic term (\ref{kineff}) and the higher-order term 
(\ref{kineff2}) in the effective low-energy theory of condensate fields
coupled to gluons. Up to mass dimension four, 
the effective theory for an $SU(3)_c$ vector $\Phi$ has, apart from
the gauge field part,
two more terms which are invariant under $SU(3)_c$ transformations 
\cite{prlett}: a mass term for the condensate field
\begin{equation} \label{masseff}
{\cal L}_{\rm eff}^{\rm mass} = {\cal M}^2 \, \Phi^\dagger \Phi\,\, ,
\end{equation}
and a quartic self interaction
\begin{equation} \label{quarticeff}
{\cal L}_{\rm eff}^{\rm int} = \lambda \, \left( \Phi^\dagger \Phi \right)^2
\,\, .
\end{equation}
Work is in progress to determine the condensate mass ${\cal M}$ 
and the coupling constant $\lambda$ \cite{dhr}.

What is the impact of these results for the solution of the gap equations?
Remember that, after taking into account the color and flavor structure,
the gap matrix in spinor space obeys the gap equation
\cite{prqcd}
\begin{equation} \label{gapequation2}
\Phi^+(K) = \frac{3}{4} \, 
g^2 \frac{T}{V} \sum_Q  \left[ \Delta^{\mu \nu}_{11}(K-Q)
- \frac{1}{9}\, \Delta^{\mu \nu}_{88} (K-Q) \right] \, \gamma_\mu \,
G_0^{-}(Q) \, \Phi^+(Q) \, G^{+}(Q)\, \gamma_\nu\,\, .
\end{equation}
Previously \cite{prlett2,prqcd,Son,SchaferWilczek}, 
the gap equation was solved using the HDL propagator for both 
$\Delta_{11}$ and $\Delta_{88}$,
\begin{equation}
\Phi^+(K) = \frac{2}{3} \, 
g^2 \frac{T}{V} \sum_Q  \Delta^{\mu \nu}_{\rm HDL}(K-Q)\, \gamma_\mu \,
G_0^{-}(Q) \, \Phi^+(Q) \, G^{+}(Q)\, \gamma_\nu\,\, ,
\end{equation}
where $\Delta_{\rm HDL}^{-1} \equiv \Delta_0^{-1} + \Pi_0$.
The integral on the right-hand side is dominated by gluons with
small momenta, $K - Q \simeq 0$. In the HDL limit,
however, static electric gluons are screened by the Debye mass,
${\Pi_0}^{00}(0) \simeq - 3\, m_g^2$, cf.\ Eq.\ (\ref{Pi00HDLstatic}). Their
contribution is therefore suppressed as compared to that of 
magnetic gluons which are not screened in the static limit,
${\Pi_0}^{ij}(0) \simeq 0$, cf.\ Eq.\ (\ref{PiijHDLstatic}).
The dominant contribution to the gap integral therefore comes from
(nearly) static magnetic gluons. A careful analysis 
\cite{prlett2,prqcd,Son,SchaferWilczek} shows that the
gluon energy is not exactly zero, but $p_0 \simeq \phi_0$, 
while the gluon momentum is $p \simeq (m_g^2 \phi_0)^{1/3}$, and thus,
in weak coupling, actually much larger than $\phi_0$. 
The coefficient $c_{\rm QCD} = 3 \pi^2/\sqrt{2}$ is determined by how 
many nearly static magnetic modes contribute, and by the precise form of the 
magnetic HDL propagator.

As shown in this paper, the gluon propagator 
in a two-flavor color superconductor is, at least in the static
limit, $p_0 =0$, and for small gluon momenta, $p \sim \phi_0$,
drastically different from the HDL propagator. 
For instance, for gluon colors 1, 2, and 3, 
which constitute the main contribution to the gap 
equation (\ref{gapequation2}), both magnetic
{\em and\/} electric modes remain unscreened.
For gluon color 8, previously unscreened static magnetic gluons attain
a Meissner mass.

In order to assess the effect of these results
on the solution of the gap equation,
one needs to solve the gap equation with the full energy and momentum
dependence of the gluon propagator in the superconducting phase, to
decide which energies and momenta constitute the dominant 
contribution to the gap integral. If gluon energy and momentum 
are much larger
than the zero-temperature gap, the impact will be rather small, because,
as was shown in Sec.\ \ref{vi}, the effect of the superconducting medium
is only a small correction of order $O(\phi_0/p)$ to the
standard HDL propagator. This might influence the prefactor
of the exponential $\exp(-c_{\rm QCD}/g)$, but not $c_{\rm QCD}$ itself.
On the other hand, if the dominant range of energies and momenta
is $p_0,\, p \sim \phi_0$, the impact could be large and might even
change $c_{\rm QCD}$.
A detailed analysis of this problem is under investigation
\cite{dsdr}.

\section*{Acknowledgements}

I thank W.\ Brown, G.\ Carter, M.\ Gyulassy, R.\ Pisarski,
K.\ Rajagopal, H.C.\ Ren, T.\ Sch\"afer, I.\ Shovkovy, and D.T.\ Son for 
discussions.
I am especially indebted to G.\ Carter, for discussions
on the ratio $\Pi_{88}(0)/\Pi_{44}(0)$, to R.\ Pisarski, 
for pointing out that ${\cal L}_{\rm eff}'$ explains the perturbative
results for the Debye and Meissner masses, and to D.T.\ Son, for
indicating the similarity of $\Pi_{88}$ with
the photon self-energy in ordinary superconductors.
My thanks go to RIKEN, BNL and the U.S.\ Dept.\ of Energy for
providing the facilities essential for the completion of this work,
and to Columbia University's Nuclear Theory Group for
continuing access to their computing facilities.
Finally, I would like to express my everlasting gratitude for the hospitality 
extended to me at Sherwood Castle, where part of this work was done.


\begin{thebibliography}{99}

\bibitem{bailinlove}
D.\ Bailin and A.\ Love, Phys.\ Rep.\ {\bf 107}, 325 (1984).

\bibitem{general}
M.\ Alford, K.\ Rajagopal, and F.\ Wilczek, Phys.\ Lett.\ {\bf B422},
247 (1998); 
R.\ Rapp, T.\ Sch\"afer, E.V.\ Shuryak, and M.\ Velkovsky,
Phys.\ Rev.\ Lett.\ {\bf 81}, 53 (1998); hep-ph/9904353;
N.\ Evans, S.D.H.\ Hsu, and M.\ Schwetz, Nucl.\ Phys.\ {\bf B551}, 
275 (1999); Phys. Lett. {\bf B449} 281, (1999);
J.\ Berges and K.\ Rajagopal, Nucl.\ Phys.\ {\bf B538}, 215 (1999);
T.\ Sch\"afer and F.\ Wilczek, Phys.\ Lett.\ {\bf B450}, 325 (1999); 
G.W.\ Carter and D.\ Diakonov, Phys.\ Rev.\ D {\bf 60}, 016004 (1999);
K.\ Langfeld and M.\ Rho, hep-ph/9811227;
M.\ Alford, J.\ Berges, and K.\ Rajagopal, hep-ph/9903502.


\bibitem{BCS} 
J.R.\ Schrieffer, {\it Theory of Superconductivity}
(New York, W.A.\ Benjamin, 1964).

\bibitem{fetter}
A.L.\ Fetter and J.D.\ Walecka, {\it Quantum Theory of Many-Particle Systems}
(McGraw--Hill, New York, 1971); A.A.\ Abrikosov, L.P.\ Gorkov, and I.E.\ 
Dzyaloshinski, {\it Methods of Quantum Field Theory in Statistical 
Physics} (Dover, New York, 1963).

\bibitem{prlett2}
R.D.\ Pisarski and D.H.\ Rischke, nucl-th/9907041 (to be published
in Physical Review D).

\bibitem{prqcd}
R.D.\ Pisarski and D.H.\ Rischke, nucl-th/9910056 (to be published
in Physical Review D).

\bibitem{prlett}
R.D.\ Pisarski and D.H.\ Rischke, Phys.\ Rev.\ Lett.\ {\bf 83}, 37 (1999).

\bibitem{Son}
D.T.\ Son, Phys.\ Rev.\ D {\bf 59}, 094019 (1999).

\bibitem{prscalar}
R.D.\ Pisarski and D.H.\ Rischke, Phys.\ Rev.\ D {\bf 60}, 094013 (1999).

\bibitem{hong}
D.K.\ Hong, hep-ph/9812510, hep-ph/9905523.


\bibitem{hongetal}
D.K.\ Hong, V.A.\ Miransky, I.A.\ Shovkovy, and L.C.R.\ Wijewardhana,
hep-ph/9906478.

\bibitem{SchaferWilczek}
T.\ Sch\"afer and F.\ Wilczek, hep-ph/9906512.

\bibitem{rockefeller}
W.E.\ Brown, J.T.\ Liu, and H.-C.\ Ren, hep-ph/9908248.

\bibitem{hsuschwetz}
S.D.H.\ Hsu and M.\ Schwetz, hep-ph/9908310.

\bibitem{LeBellac}
M.\ Le Bellac, {\it Thermal Field Theory}
(Cambridge, Cambridge University Press, 1996).

\bibitem{finitedens}
J.-P.\ Blaizot and J.-Y.\ Ollitrault, Phys.\ Rev.\ D {\bf 48}, 1390 (1993);
H.\ Vija and M.H.\ Thoma, Phys.\ Lett.\ {\bf B342}, 212 (1995);
C.\ Manuel, Phys.\ Rev.\ D {\bf 53}, 5866 (1996).

\bibitem{ehhs}
N.\ Evans, J.\ Hormuzdiar, S.D.H.\ Hsu, and M.\ Schwetz,
hep-ph/9910313.

\bibitem{parity}
R.D.\ Pisarski and D.H.\ Rischke, nucl-th/9906050.

\bibitem{rp}
R.D.\ Pisarski, nucl-th/9912070.

\bibitem{shovkovy}
The existence of such a term was pointed out by
I.A.\ Shovkovy.

\bibitem{sonstephanov}
D.T.\ Son and M.A.\ Stephanov, hep-ph/9910491.

\bibitem{dhr}
D.H.\ Rischke, (work in progress).

\bibitem{dsdr}
D.H.\ Rischke and D.T.\ Son, (work in progress).









\end{thebibliography}
\end{document}